\documentclass[]{aa}

\usepackage{graphicx}

\usepackage{amsmath}
\usepackage{amssymb}
\usepackage{multirow}
\usepackage{arydshln}

\usepackage{lipsum}
\usepackage{soul}
\usepackage[normalem]{ulem}

\usepackage{natbib}
\usepackage{hyperref}

\usepackage{color, colortbl}
\definecolor{medgray}{gray}{0.7}
\definecolor{lightgray}{gray}{0.92}

\begin{document}

\title{The orbit of HD 142527 B is too compact to explain many of the disc features}

\authorrunning{M. Nowak, S. Rowther, et al.}

\author{M.~Nowak\inst{\ref{ioa},\ref{kavli}}, S. Rowther\inst{\ref{Warwick1},\ref{Warwick2},\ref{Leicester}}, S. Lacour\inst{\ref{lesia},\ref{eso}}, F. Meru\inst{\ref{Warwick1},\ref{Warwick2}}, R. Nealon\inst{\ref{Warwick1},\ref{Warwick2}}, D. J. Price\inst{\ref{monash}}}

\institute{
  Institute of Astronomy, University of Cambridge, Madingley Road, Cambridge CB3 0HA, United Kingdom
  \email{mcn35@cam.ac.uk} 
  \label{ioa} \and
  Kavli Institute for Cosmology, University of Cambridge, Madingley Road, Cambridge CB3 0HA, UK
  \label{kavli}  \and
  Centre for Exoplanets and Habitability, University of Warwick, Coventry CV4 7AL, UK
  \label{Warwick1} \and
  Department of Physics, University of Warwick, Coventry CV4 7AL, UK
  \label{Warwick2} \and
  School of Physics and Astronomy, University of Leicester, Leicester LE1 7RH, UK
  \label{Leicester} \and
  LESIA, Observatoire de Paris, PSL, CNRS, Sorbonne Université, Université de Paris, 5 place Janssen, 92195 Meudon, France
  \label{lesia} \and  
  European Southern Observatory, Karl-Schwarzschild-Stra{\ss}e 2, 85748 Garching, Germany
  \label{eso} \and
  School of Physics and Astronomy, Monash University, Clayton Vic 3800, Australia
  \label{monash}
}

\newcommand{\LEt}[1]{\textcolor{red}{#1}}

\date{\today}
  \abstract
  {HD~142527~A is a young and massive Herbig Ae/Be star surrounded by a highly structured disc. The disc shows numerous morphological structures, such as spiral arms, a horseshoe region of dust emission, a set of shadows cast by an inner disc on the outer disc, and a large cavity extending from $\simeq{}30\,\mathrm{au}$ to $\simeq{}130\,\mathrm{au}$. HD~142527~A also has a lower mass companion, HD~142527~B ($M=0.13\pm{}0.03\,M_\odot$), which is thought to be responsible for most of the structures observed in the surrounding disc.}
{We aim to fully constrain the orbit of HD~142527~B and determine whether the binary alone is truly responsible for the observed morphology of the HD~142527 disc.}
{We gathered VLTI/GRAVITY observations of HD~142527, either from our own programmes or from the ESO archive. We used this inhomogeneous set of data to extract a total of seven  high-precision measurements of the relative astrometry between HD~142527~A and B, spread from mid-2017 to early 2021. Combined with what is available in the literature, these new measurements offer a total of 9 yr of astrometric monitoring on HD~142527. We used orbit fitting tools to determine the orbital parameters of HD~142527~B, and used them as inputs for a 3D hydrodynamical model of the disc to determine whether or not the binary is able to create the structures observed in the disc.}
{Our VLTI/GRAVITY astrometry gives excellent constraints on the orbit of HD~142527~B. We show that the secondary is following an orbit of semi-major axis $a=10.80\pm0.22\,\mathrm{au}$, with moderate eccentricity ($e=0.47\pm{}0.01$), and has recently passed its periapsis ($\tau=2020.42$). With such a compact orbit, we show that HD~142527~B can only generate a gap and spiral arms of $\sim{}30\,\mathrm{au}$ in the disc, which is much smaller than what is revealed by observations.}
{Even from a theoretical standpoint, the observed cavity size of $\sim{}100\,\mathrm{au}$ far exceeds even the most generous predictions for a companion like HD~142527~B on such a compact orbit. Thus, we conclude that the low-mass companion cannot be solely responsible for the observed morphology of the disc surrounding the system.}


\maketitle

\section{Introduction}
\label{sec:intro}

HD~142527~A is a $5.0\pm{}1.5\,\mathrm{Myr}$ old Herbig Ae/Be star of type F6-IIIV, with a mass of $2.0\pm{}0.3\,M_\odot$ \citep{Mendigutia2014}, located at a distance of $159.3\pm{}0.7\,\mathrm{pc}$ \citep{GaiaCollaboration2021}. The system has been extensively studied for its spectacular circumstellar disc, which harbours several enigmatic morphological features. Observations in the near-infrared with Subaru/CIAO have revealed the presence of a large spiral arm at $>100\,\mathrm{au}$ \citep{Fukagawa2006}, with a counterpart in CO later detected with the Atacama Large Millimetre/submillimetre Array \citep[ALMA,][]{Christiaens2014}. Smaller spiral arms were also detected in the near-IR \citep{Casassus2012, Canovas2013, Avenhaus2014, Avenhaus2017}, as well as other larger ones, extending up to $\simeq{}500\,\mathrm{au}$ \citep{Christiaens2014}.

The main feature of the HD~142527 disc is arguably its large central cavity, which extends from $\simeq{}30$ to $\simeq{}130\,\mathrm{au}$. Its presence was first suggested by \cite{Fukagawa2006}, before being confirmed with ALMA by \cite{Casassus2013}, who also detected a horseshoe structure of continuum emission surrounding this central cavity.

Early observations with the MIDI instrument at the Very Large Telescope Interferometer (VLTI) revealed the presence of an inner disc, which has since been detected in visible light with SPHERE/Zimpol \citep{Avenhaus2017}. This inner disc is thought to be inclined at $\simeq{}70\,\mathrm{deg}$ relative to the outer disc, thus casting a narrow shadow on it, and explaining some of the azimuthal asymmetries observed in the near-IR \citep{Casassus2012, Marino2015}.

Its highly structured morphology, eccentric outer disc, and high fraction of crystalline silicates \citep{vanBoekel2004, Avenhaus2017}, have made HD~142527 a prime taget for planet searches. The low-mass companion HD~142527~B was discovered by \cite{Biller2012} using the Sparse Aperture Masking (SAM) mode of SPHERE/NACO, and this new object quickly became the prime suspect to explain the complex disc structures. \cite{Lacour2016} used the SAM mode of SPHERE/NACO and Gemini/GPI to further characterise the companion and its orbit. They showed that HD~142527~B was likely an M dwarf with a mass of $0.13\pm{}0.03\,M_\odot{}$ and a temperature of $T=3000\pm{}100\,\mathrm{K}$ (i.e. with a contrast ratio of $\simeq{}1/75$ and a mass ratio of $\simeq{}1/10$ with respect to the primary). Their astrometric measurements covered epochs ranging from early 2012 to  mid-2014, and suggested two possible families of orbits: one that had passed periapsis, and another  approaching periapsis. Both families had moderate eccentricities ($e\simeq{}0.5$), and semi-major axes ranging from $70$ to $260\,\mathrm{au}$.

The first comprehensive study linking the low-mass companion HD~142527~B to the morphological structure of the disc was conducted by \cite{Price2018}. With the use of a series of 3D hydrodynamical models of HD~142527, they showed that a low-mass companion, with characteristics compatible with that reported by \cite{Lacour2016}, could explain most of the observed structures of the disc. Their study also strongly favoured the family of orbit approaching periapsis, as only this one was compatible with the disc features. \cite{Claudi2019} further refined the orbital parameters of the binary with a set of new astrometric measurements with SPHERE, and reinforced the suggestion of \cite{Price2018} that HD~142527~B was approaching its periapsis.

More recently, \cite{Balmer2022} has presented the most extensive set of astrometric measurements to date on HD~142527~B, with epochs ranging from early 2012 to mid-2018. They were able to give good constraints on most of the orbital parameters, and concluded that the low-mass companion was indeed approaching its periapsis over this period. They reported a semi-major axis of $\simeq{}15\,\mathrm{au}$, an eccentricity of $\simeq{}0.3$, and an orbit close to perpendicular to the outer disc (mutual inclination of $\simeq{}80\,\mathrm{deg}$). They concluded that this high mutual inclination supported the idea that the binary was sculpting the disc, as the nearly perpendicular orbit was one of their main assumptions needed by \cite{Price2018} to explain the disc features.

We note, however, that both the semi-major axis and the eccentricity derived by \citeauthor{Balmer2022} ($a\simeq{}15\,\mathrm{au}$ and $e\simeq{}0.3$) are significantly smaller than the values used by \citeauthor{Price2018} ($a\simeq{}30\,\mathrm{au}$ and $e\simeq{}$0.6-0.7). As the lower semi-major axis and eccentricity are expected to result in different disc structures, this raises the question of whether the companion can really explain the complex morphological structures observed in the disc.

In this work we attempt to answer this question with a set of VLTI/GRAVITY observations gathered both from the ESO archive and from our own observing  programmes. These long-baseline interferometric obervations offer a significant gain in terms of astrometric precision compared to single-dish observations, resulting in better constraints on the orbital parameters of the binary.

The paper is structured as follows. In Section~\ref{sec:observations} we introduce our observations, and in Section~\ref{sec:data_reduction} we give some details on how we extracted astrometric measurements from these data, which were obtained in a variety of modes and configurations. In Section~\ref{sec:orbit} we focus on the orbit fit, and present our new constraints on the orbital parameters of the binary, and on its mutual inclination with the inner and outer discs. We use these orbital parameters as an input in a set of 3D hydrodynamical simulations of the disc, and discuss the results in Section~\ref{sec:disc}. Finally, we give our conclusions in Section~\ref{sec:conclusion}. 


\section{Observations}
\label{sec:observations}

We gathered a number of observations of HD~142527, obtained with VLTI/GRAVITY between 2017 and 2021. The observations are spread over ten different nights. Some of them are part of regular programmes targeting HD~142527, whereas others were acquired using HD~142527 either as a backup target in case of bad weather, or as a target for technical tests with the instrument. As a consequence, the set of observations presented in this paper is inhomogeneous, with a mixture of VLTI configurations (auxiliary telescopes, ATs, or  unit telescopes, UTs), spectral resolutions (medium or high), polarisation settings (combined or split), and observing modes (dual-field or single-field). The full observing log is presented in Table~\ref{tab:obslog}.

In principle, single-field observations with GRAVITY are used when the field of view of the fibre ($\sim{}300\,\mathrm{mas}$ in diameter at half-maximum with the ATs, $\sim{}60\,\mathrm{mas}$ with the UTs) is larger than the separation of the observed binary. In this case the light from both components is injected into the fibre, and the data are analysed using the squared visibilities and closure phases formalism (see Section~\ref{sec:data_reduction}). However, when the separation is larger than the field of view of the fibre, the instrument is set to dual-field mode, where the two components can be observed sequentially by moving the fibre back and forth between the two. This strategy, used in particular for exoplanet observations, is described in \citep{GravityCollaboration2019, GravityCollaboration2020}.

In the case of HD~142527, the separation is typically $\sim{}50\,\mathrm{mas}$, and the observations were carried out in single-field when observing with the ATs, and in dual-field when observing with the UTs (see Table~\ref{tab:obslog}). However, we found that in all the dual-field observations, when the fibre is centred on   HD~142527~B, it also collects a significant amount of flux from HD~142527~A, due to the small separation and high contrast ratio ($\sim{}1/75$ in K-band) between the two components. As a consequence, we decided to treat these observations as single-field rather than dual-field, and to discard the dual-field data in which the fibre is centred on the star, as they are too largely dominated by HD~142527~A. We report only the observations obtained at the location of HD~142527~B in Table~\ref{tab:obslog}. Finally, when the data obtained are in split polarisation, we treat the polarisations separately, as if they were independent exposures.

\begin{table*}
    \centering
    \footnotesize
    \begin{tabular}{c l l c c c c c}
        \hline
        \hline
         Epoch & Configuration$^{1}$ & Mode$^{2}$ & $N_\mathrm{exp}\times{}N_\mathrm{DIT}\times{}\mathrm{DIT}$ & Airmass & Seeing (") & Programme \\
         \hline
         2017-03-19$^\dagger$ & ATs Astro & Single,High,Combined & $7\times{}10\times{}30\,\mathrm{s}$ & 1.05-1.07 & 0.6-0.81 & 098.D-0488(A) \\
         \hline
         \multirow{2}{*}{2017-05-03} & \multirow{2}{*}{ATs Large} & Single,Medium,Combined & $2\times{}30\times{}10\,\mathrm{s}$ & 1.22-1.25 & 0.33-0.43 & \multirow{2}{*}{Technical} \\
         \cdashline{3-6}[0.5pt/3pt]
          & & Single,Medium,Split & $3\times{}20\times{}30\,\mathrm{s}$ & 1.14-1.67 & 0.41-0.65 & \\
         \hline
         \multirow{3}{*}{2018-02-17} & \multirow{3}{*}{ATs Large} & \multirow{2}{*}{Single,Medium,Combined} & $2\times{}30\times{}0.3\,\mathrm{s}$ & \multirow{2}{*}{1.86-2.28} & \multirow{2}{*}{0.48-0.67} & \multirow{3}{*}{60.A-9800(U)} \\
         & & & $4\times{}30\times{}1\,\mathrm{s}$ & & & \\
         \cdashline{3-6}[0.5pt/3pt]
          &  & Single,High,Combined & $2\times{}20\times{}10\,\mathrm{s}$ & 1.71-1.78 & 0.46-0.48 &  \\
         \hline
         \multirow{3}{*}{2018-07-25} & \multirow{3}{*}{UTs} & Dual,High,Split & $2\times{}15\times{}30\,\mathrm{s}$ & 1.05-1.05 & 1.10-1.54 & \multirow{3}{*}{60.A-9800(U)} \\
         \cdashline{3-6}[0.5pt/3pt]
         & & \multirow{2}{*}{Dual,Medium,Split} & $2\times{}20\times{}10\,\mathrm{s}$ & \multirow{2}{*}{1.06-1.07} & \multirow{2}{*}{0.96-1.39} & \\
         & & & $2\times{}10\times{}10\,\mathrm{s}$ & & & \\         
         \hline
         2018-07-28 & UTs & Dual,Medium,Split & $2\times{}15\times{}10\,\mathrm{s}$ & 1.09-1.11 & 1.13-1.46 & 60.A-9800(U) \\
         \hline
         2019-04-27$^\dagger$ & ATs Medium & Single,Medium,Split & $2\times{}10\times{}30~\mathrm{s}$ & 1.10-1.13 & 0.66-0.73 & 60.A-9132(A)\\         
         \hline
         2019-04-28 & ATs Medium & Single,Medium,Split & $2\times{}10\times{}30\,\mathrm{s}$ & 1.12-1.15 & 0.69-0.88 & 60.A-9132(A) \\         
         \hline
         2019-05-01 & ATs Medium & Single,Medium,Split & $4\times{}10\times{}30\,\mathrm{s}$ & 1.06-1.10 & 1.21-2.06 & 60.A-9132(A) \\         
         \hline
         2019-05-10$^\dagger$ & ATs Small & Single,Medium,Split & $4\times{}10\times{}30\,\mathrm{s}$ & 1.06-1.25 & 0.64-1.09 & 60.A-9132(A) \\         
         \hline
         2021-01-08 & UTs & Dual,Medium,Combined & $8\times{}32\times{}1\,\mathrm{s}$ & 1.70-1.80 & 0.59-0.81 & 1104.C-0651(C) \\         
         \hline
    \end{tabular}
    \caption{Observing log presenting all   data acquired with VLTI/GRAVITY on HD~142527 and analysed in this paper. \newline
      $^1$ ATs: auxiliary telescopes; UTs: unit telescopes. A description of the AT  configurations is available on the \href{https://www.eso.org/sci/facilities/paranal/telescopes/vlti/configuration/P105.html}{ESO website}. \newline
      $^2$ Single: single-field mode; Dual: dual-field mode; Medium or high: spectral resolution; Combined or split: polarisation setting. \newline
      $^\dagger$ We were not able to extract an astrometry from these datasets (see Section~\ref{sec:data_reduction}). They are listed here only for completeness, as they are available on the ESO archive.
    }
    \label{tab:obslog}  
\end{table*}

\section{Data reduction}
\label{sec:data_reduction}

All the observations were initially reduced using the GRAVITY pipeline \citep{Lapeyrere2014}. We generated scivis files, which were averaged over all DITs of each exposure, and which contained the squared modulus of the interferometric visibilities, $|V^2|$, and the closure phases, $\mathcal{C}_\phi$, with the associated error bars. The squared visibilities and the closure phases both encode information about the separation of the two components of the binary, and we treated them as independent quantities.

The coherent flux received at time $t$, along the baseline $b$, and at wavelength $\lambda$ can be written
\begin{equation}
  \mathcal{F}(t, b, \lambda) = S_A(\lambda) + S_B(\lambda)\times{}e^{-j\frac{2\pi}{\lambda}\mathrm{OPD}(t, b, \lambda)}
  \label{eq:1}
,\end{equation}
\noindent{}where $S_A$ and $S_B$ are the spectra of the two components of the binary, and $\mathrm{OPD}$ is the optical path difference between the two components of the binary along baseline $b$.

\noindent{}Using the coordinates of the baseline in the \emph{uv}-plane $b=(u, v)$, and the separation vector $(\Delta\alpha, \Delta\delta)$ of the binary in RA/DEC, the OPD can be written as
\begin{equation}
  \mathrm{OPD} = \Delta{}\alpha\times{}u+\Delta{}\delta{}\times{}v
  \label{eq:2}    
.\end{equation}

\noindent{}Equation~\ref{eq:1} thus becomes
\begin{equation}
  \mathcal{F} = S_A(\lambda) + S_B(\lambda)\times{}e^{-j\frac{2\pi}{\lambda}\left(\Delta{}\alpha\times{}u+\Delta{}\delta{}\times{}v\right)}
  \label{eq:3}  
.\end{equation}

Equation~\ref{eq:3} presents an idealised case, as it does not take into account the phase distortions introduced by the atmosphere. The Fringe Tracker of GRAVITY \citep{Lacour2019} corrects this atmospheric phase distortion up to a near-constant term, which affects the telescopes individually and can thus be represented by a set of four terms: $\phi_{T_1}, \phi_{T_2}, \phi_{T_3}, \phi_{T_4}$. For a baseline $b$ linking telescope $k$ to telescope $l$, the post fringe-tracker atmospheric phase distortion is $\Delta\phi = \phi_{T_l} - \phi_{T_k}$. This term affects both components of the binary in the same way, and Equation~\ref{eq:3} thus becomes
\begin{equation}
  \mathcal{F}_{T_kT_l} = \left[S_A + S_B\times{}e^{-j\frac{2\pi}{\lambda}\left(\Delta{}\alpha\times{}u+\Delta{}\delta{}\times{}v\right)}\right]e^{j\left(\phi_{T_l} - \phi_{T_k}\right)}
  \label{eq:4}  
,\end{equation}
\noindent{}which shows that Equation~\ref{eq:4} encodes information about the separation of the binary in two forms: in the amplitude of the observed visibility and in the phase.

The flux received at each telescope $T$ of the array is simply $F_T = S_A+S_B$, and the interferometric visibility is therefore given by
\begin{align*}
  V_{T_kT_l} &= \frac{\mathcal{F}_{T_kT_l}}{\sqrt{F_{T_k}F_{T_l}}} \\
  & = \frac{{\left[1 + Ce^{-j\frac{2\pi}{\lambda}\left(\Delta{}\alpha\times{}u+\Delta{}\delta{}\times{}v\right)}\right]e^{j\left(\phi_{T_l} - \phi_{T_k}\right)}}}{1+C},
\end{align*}
\noindent{}where $C = S_B/S_A$ is the contrast of the binary.

The squared modulus of the visibility for a baseline, $T_kT_l$, and the closure phase for a triangle, $T_kT_lT_m$, are defined by
\begin{align*}
    |V_{T_kT_l}|^2 &= V_{T_kT_l}V^*_{T_kT_l} \\
    \mathcal{C}_{T_kT_lT_m} &= \arg{(V_{T_kT_l}V_{T_lT_m}V^*_{T_mT_k})}.
\end{align*}
\noindent{}Using Equation~\ref{eq:4} in combination with these definitions, we see that the two quantities are independent of the residual atmospheric distortion term, and are given by
\begin{align}
    |V_{T_kT_l}|^2 &= \frac{1+C^2+2C\cos{(\Phi_{T_kT_l})}}{(1+C)^2} \label{eq:5},\\
    \mathcal{C}_{T_kT_lT_m} &= \arg{(1+Ce^{-j\Phi_{T_kT_l}})}+\arg{(1+Ce^{-j\Phi_{T_lT_m}})} \nonumber\\ & \qquad -\arg{(1+Ce^{-j\Phi_{T_kT_m}}),}
    \label{eq:6}
\end{align}

\noindent{}where $\Phi_{T_iT_j}$ is the phase corresponding the baseline $(u_{ij}, v_{ij})$ linking telescope $T_i$ to $T_j$:
\begin{align*}
  \Phi_{T_iT_j} = \frac{2\pi}{\lambda}\left(\Delta{}\alpha\times{}u_{ij}+\Delta{}\delta{}\times{}v_{ij}\right).
\end{align*}

In principle, since the $(u,v)$ coordinates of the different baselines are known, Equations~\ref{eq:5} and \ref{eq:6} can be used to model the GRAVITY data with only three parameters: the binary separation vector $(\Delta{}\alpha, \Delta{}\delta)$, and the contrast ratio $C$ (under assumption of a constant contrast along the GRAVITY wavelength range, from $\sim{}1.9\,\mu\mathrm{m}$ to $2.4\,\mu\mathrm{m}$).

In the absence of proper calibration, however, both the squared visibilities and the closure phases are affected by additional low-order terms, which we modelled with the use of an additional polynomial in wavelength  of order 2. In Equation~\ref{eq:5} the constant terms can be absorbed in this low-order polynomial. Denoting $a_0, a_1, a_2$ the coefficients of the additional polynomial term, and using $\beta = 2C/(1+C)^2$ as a parameter instead of $C$, our model for the squared visibility becomes linear in all parameters but $(\Delta\alpha, \Delta\delta)$. Our model for the closure phase remains non-linear in three parameters ($(\Delta\alpha, \Delta\delta)$, and $C$) and is written as
\begin{equation}
  |V_{T_kT_l}|^2_\mathrm{model} = a_0+a_1\lambda+a_2\lambda^2+\beta\cos{[\Phi_{T_kT_l}(\Delta\alpha, \Delta\delta)]} \label{eq:7}
,\end{equation}
\begin{align}
    \mathcal{C}_{T_kT_lT_m, \mathrm{model}} & = a_0+a_1\lambda+a_2\lambda^2+\arg{(1+Ce^{-j\Phi_{T_kT_l}})} \nonumber{} \\ & \qquad +\arg{(1+Ce^{-j\Phi_{T_lT_m}})} \nonumber \\ & \qquad -\arg{(1+Ce^{-j\Phi_{T_kT_m}}).}
    \label{eq:8}
\end{align}

The GRAVITY pipeline reports the squared visibilities and the closure phases separately for each baseline and wavelength, with the associated error bars. Therefore, we were able to define two $\chi^2$ quantities for each file of each epoch of our single-field observations (see Table~\ref{tab:obslog}) by summing the $\chi^2$ obtained using the models defined in Equations~\ref{eq:7} and \ref{eq:8} over the six baselines (for $|V|^2$) and four triangles (for $\mathcal{C}$):
\begin{align*}
  \chi_{|V|^2}^2 &= \sum_{T_iT_j, \lambda}\left(\frac{|V|^2_\mathrm{data}(T_iT_j, \lambda) - |V|^2_\mathrm{model}(T_iT_j, \lambda)}{\sigma_{|V^2|}(T_iT_j, \lambda)}\right)^2, \\
  \chi_{\mathcal{C}}^2 &= \sum_{T_iT_jT_k, \lambda}\left(\frac{\mathcal{C}_\mathrm{data}(T_iT_jT_k, \lambda) - \mathcal{C}_\mathrm{model}(T_iT_jT_k, \lambda)}{\sigma_\mathcal{C}(T_iT_jT_k, \lambda)}\right)^2.
\end{align*}

From this we calculated, for each file, a set of two $\chi^2$ maps as a function of $\Delta\alpha, \Delta\delta$, by minimising the two $\chi^2$ over all the nuisance parameters. This represents a minimisation over 19 linear parameters for the squared visibilities (a set of $a_0, a_1, a_2$ for each of the six baselines, plus a value for $\beta$), and 12 linear parameters ($a_0, a_1, a_2$ for each of the four triangles) plus one non-linear parameter (the contrast value $C$) for the closure phases: 
\begin{align}
  \chi^2_{|V|^2,\mathrm{map}}({\Delta{}\alpha}, {\Delta{}\delta}) &= \mathrm{min}_{a_0, a_1, a_2, \beta}\left(\chi^2_{|V|^2}\right) \label{eq:9},\\
  \chi^2_{\mathcal{C},\mathrm{map}}({\Delta{}\alpha}, {\Delta{}\delta}) &= \mathrm{min}_{a_0, a_1, a_2, C}\left(\chi^2_{\mathcal{C}}\right).  \label{eq:10}
\end{align}

In Equation~\ref{eq:9} we used the closed-form solution of the minimisation problem since all parameters are linear. For Equation~\ref{eq:10} we used the closed-form solution to minimise over the linear parameters, and the Powel algorithm to perform the minimisation over the non-linear parameter $C$.

\begin{figure*}
    \centering
    \includegraphics[width=0.99\linewidth]{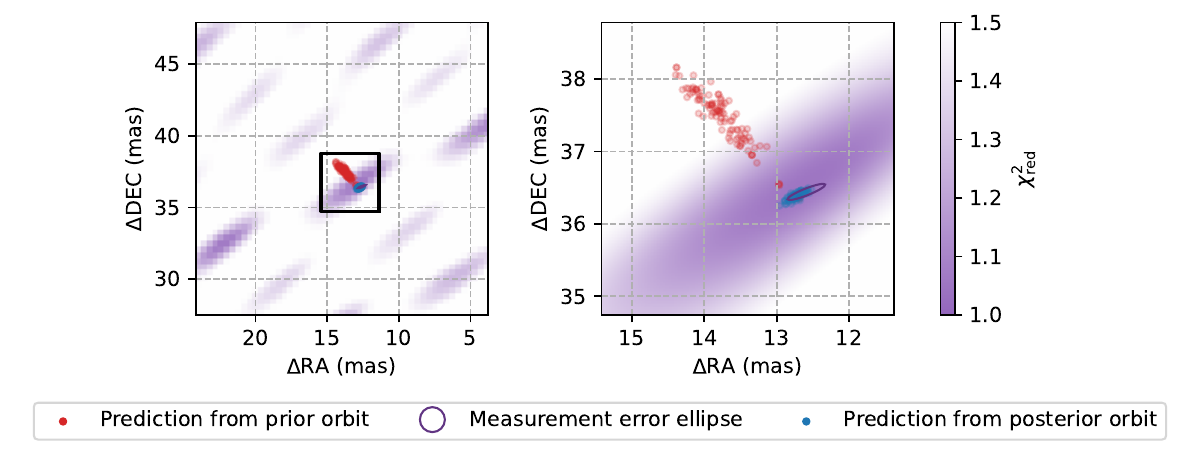}
    \caption{Typical example of a $\chi^2$ map obtained when reducing our data (in this case, for the observation on 2019 April 28). The left panel shows the structure of the $\chi^2$ map on a large scale, where the multiple minimums (in purple) are clearly visible. The right panel is a zoomed-in version. In each panel the red dots are generated by performing an orbit fit with all the astrometric measurement data available (including all GRAVITY epochs up to 2019 April 28), and using the posterior distributions of orbital parameters to calculate the estimated position at epoch 2019 April 28. The purple ellipse corresponds to the 3$\sigma$ contour for the astrometric solution. The blue dots are obtained in a similar way, but using a new orbit fit that includes the astrometry extracted from this epoch. In other words, the red dots represents the prediction using the {a priori} orbital solution, the purple ellipse represents the measurement obtained at this epoch, and the blue dots represent the prediction using the {a posteriori} updated orbital solution.}
    \label{fig:chi2maps}
\end{figure*}

An example of a typical $\chi^2$ map is given in Figure~\ref{fig:chi2maps}. It shows a structure with multiple adjacent minimums, which is characteristic of interferometric observations with a limited number of baselines. To identify the correct minimum, we make use of prior information on the orbit of the binary. We analysed our data sequentially, one epoch at a time, starting from the first one. For each epoch $t$, we took all the astrometric data available in the literature (see Table~\ref{tab:astrometry}), to which we added our astrometric measurements obtained on the previous GRAVITY epochs (with $t'<t$). We then used \verb|orbitize!| \citep{Blunt2020} to perform an orbit fit for the binary. We gathered a total of 100\,000 posterior values for the orbital parameters of the binary, which we used to calculate the expected astrometry at epoch $t$ (see Figure~\ref{fig:chi2maps}). For all of our epochs, this {a priori} estimate was good enough to unambiguously determine the correct minimum in our $\chi^2$ maps.

Once we had identified the rough location of the correct minimum in each map, we took the location of the closest minimum to be our best estimate of the astrometry, making sure that the $\chi^2$ is calculated with a resolution that is much smaller than our final error bars.

For each epoch, we therefore obtained a set a $2\times{}N_\mathrm{exp}$ estimates for the astrometry of the binary. Our final best estimate for this epoch was taken as the mean of all these measurements, and the error ellipse was calculated using an empirical estimate of the covariance matrix.

We were not able to obtain a proper astrometric measurement for three of our ten epochs. All these difficult datasets were obtained in single-field with the ATs.

The data obtained on March 19, 2017, appeared corrupted even on simple visual inspection, and were very badly fitted by our models, which makes us suspect a fringe-tracker problem during the observations. On April 27, 2019, the baselines were aligned in a very unfortunate way, such that only one of them produces a signal significantly different from the second-order polynomial used to detrend the data. With only a single useful baseline, a measurement of the astrometry is not possible, and thus the data were discarded. Finally, on May 10, 2019, the observations were performed using the small set-up with the ATs, which was unable to produce a signal significantly different from our second-order detrending polynomial due to the small separation of the target. The results obtained for the three epochs of dual-field UT observations, and the four epochs of single-field AT observations for which we were able to obtain an astrometric measurement are given in Table~\ref{tab:astrometry}.

\begin{table*}
    \centering
    \footnotesize
    \begin{tabular}{c c c c c c c l l}
        \hline
        \hline
        \multirow{2}{*}{Epoch} & \multirow{2}{*}{MJD} & Separation$^*$ & Position Angle$^*$ & $\Delta{}\mathrm{RA}$ & $\Delta{}\mathrm{DEC}$ & $\rho^{**}$ & \multirow{2}{*}{Instrument} & \multirow{2}{*}{Reference$^{***}$} \\
        & & (mas) & (deg) & (mas) & (mas) & - \\
        \hline
2012-03-11 & 55997 & $89.70\pm 2.60$ & $133.10\pm 1.90$ & $65.50\pm 10.79$ & $-61.29\pm 11.48$ & 0.92 & NACO & Bl12 \\
\hline
2013-03-17 & 56368 & $82.00\pm 2.10$ & $126.30\pm 1.60$ & $66.09\pm 2.16$ & $-48.55\pm 2.22$ & 0.08 & NACO & L16 \\
\hline
2013-04-11 & 56393 & $82.50\pm 1.01$ & $127.98\pm 0.46$ & $65.03\pm 0.89$ & $-50.77\pm 0.81$ & -0.39 & VisAO & Ba22 \\
\hline
2013-07-14 & 56487 & $82.50\pm 1.10$ & $123.80\pm 1.20$ & $68.56\pm 1.32$ & $-45.89\pm 1.56$ & 0.40 & NACO & L16 \\
\hline
2014-04-08 & 56755 & $77.73\pm 1.63$ & $117.87\pm 0.96$ & $68.71\pm 1.56$ & $-36.34\pm 1.39$ & -0.18 & VisAO & Ba22 \\
\hline
2014-04-25 & 56772 & $88.25\pm 10.10$ & $123.00\pm 9.20$ & $74.01\pm 11.44$ & $-48.06\pm 13.03$ & 0.30 & GPI & R14 \\
\hline
2014-05-10 & 56787 & $78.00\pm 1.80$ & $119.10\pm 1.00$ & $68.15\pm 1.71$ & $-37.93\pm 1.48$ & -0.23 & SINFONI & Ch18 \\
\hline
2014-05-12 & 56789 & $77.20\pm 0.60$ & $116.60\pm 0.50$ & $69.03\pm 0.62$ & $-34.57\pm 0.66$ & 0.09 & GPI & L16 \\
\hline
2015-05-13 & 57155 & - & - & $65.00\pm 2.00$ & $-24.00\pm 2.00$ & - & SPHERE & Cl18 \\
\hline
2015-05-15 & 57157 & $70.16\pm 1.19$ & $110.56\pm 0.80$ & $65.69\pm 1.17$ & $-24.64\pm 1.01$ & -0.12 & VisAO & Ba22 \\
\hline
2015-05-16 & 57158 & $72.19\pm 2.02$ & $107.84\pm 0.97$ & $68.72\pm 1.95$ & $-22.12\pm 1.32$ & -0.29 & VisAO & Ba22 \\
\hline
2015-05-18 & 57160 & $70.00\pm 1.35$ & $110.12\pm 0.72$ & $65.73\pm 1.30$ & $-24.08\pm 0.95$ & -0.28 & VisAO & Ba22 \\
\hline
2015-07-03 & 57206 & - & - & $62.90\pm 0.40$ & $-18.30\pm 0.50$ & - & SPHERE/SAM & Cl18 \\
\hline
2016-03-26 & 57473 & - & - & $60.00\pm 2.00$ & $-7.00\pm 2.00$ & - & SPHERE & Cl18 \\
\hline
2016-03-31 & 57478 & $62.80\pm 2.70$ & $98.70\pm 1.80$ & $62.08\pm 2.69$ & $-9.50\pm 2.00$ & -0.10 & ZIMPOL & Cu19 \\
\hline
2016-06-13 & 57552 & - & - & $61.00\pm 2.00$ & $-7.00\pm 2.00$ & - & SPHERE & Cl18 \\
\hline
\rowcolor{lightgray}
2017-05-03 & 57876 & - & - & $49.67\pm 0.04$ & $11.33\pm 0.08$ & 0.50 & GRAVITY & This work \\
\hline
2017-05-16 & 57889 & - & - & $47.40\pm 0.50$ & $10.30\pm 0.20$ & - & SPHERE/SAM & Cl18 \\
\hline
\rowcolor{lightgray}
2018-02-17 & 58166 & - & - & $38.07\pm 0.50$ & $23.49\pm 0.28$ & -0.79 & GRAVITY & This work \\
\hline
2018-04-14 & 58222 & - & - & $36.00\pm 1.00$ & $25.00\pm 1.00$ & - & SPHERE & Cl18 \\
\hline
2018-04-27 & 58235 & $43.69\pm 2.01$ & $58.28\pm 1.76$ & $37.16\pm 1.85$ & $22.97\pm 1.55$ & 0.35 & VisAO & Ba22 \\
\hline
\rowcolor{lightgray}
2018-07-25 & 58324 & - & - & $29.70\pm 0.04$ & $29.65\pm 0.08$ & -0.43 & GRAVITY & This work \\
\hline
\rowcolor{lightgray}
2018-07-28 & 58327 & - & - & $29.48\pm 0.02$ & $29.51\pm 0.08$ & -0.30 & GRAVITY & This work \\
\hline
\rowcolor{lightgray}
2019-04-28 & 58601 & - & - & $12.59\pm 0.16$ & $36.44\pm 0.10$ & -0.93 & GRAVITY & This work \\
\hline
\rowcolor{lightgray}
2019-05-01 & 58604 & - & - & $12.48\pm 0.15$ & $36.53\pm 0.09$ & -0.90 & GRAVITY & This work \\
\hline
\rowcolor{lightgray}
2021-01-08 & 59222 & - & - & $-27.77\pm 0.22$ & $23.98\pm 0.05$ & -0.49 & GRAVITY & This work \\
\hline
    \end{tabular}
    \caption{Astrometry on HD~142527 B. \newline
      $^*$ Different measurements made with different techniques and instruments are reported on two different bases: separation and position angle, or right-ascension (RA) and declination (DEC). For convenience, we have converted all values reported in separation/position angle to RA/DEC using an MCMC approach, but these values are treated on their original bases during our orbit fits. \newline
      $^{**}$ When available, the correlation coefficient $\rho$ is reported in addition to the error along the RA and DEC axes. \newline
      $^{***}$ Sources: Ba22: Balmer 2022; L16: Lacour 16; Bl12: Biller 2012; R14: Radius 2014; Ch18: Christiaens 2018; Cl19: Claudi 2019; Cu19: Cugno 2019.}
    \label{tab:astrometry}
\end{table*}

\section{The orbit of HD~142527}
\label{sec:orbit}

\subsection{Orbital parameters}

Including both AT and UT observations, we report in this work a set of seven new astrometric measurements of HD~142527~B, with typical precisions ranging from a few $100\,\mu\mathrm{as}$ down to $\simeq{}10\,\mu\mathrm{as}$, obtained between 2017 and early 2021. To this we added a total of 19 epochs compiled from the literature to create a dataset that covers almost one half of the orbit of the binary, from 2012 to 2021.

We used \verb|orbitize!| \citep{Blunt2020} to fit the relative motion of HD~142527~B around HD~142527~A, and to derive estimates of its orbital parameters. We used the parallel-tempering affine-invariant \citep{Foreman-Mackey2013, Vousden2016}  Markov chain Monte Carlo (MCMC) sampler available in \verb|orbitize!| to determine the posterior distributions of the eight orbital parameters reported in Table~\ref{tab:parameters}. Except for the total mass of the system, $M_\mathrm{tot}$, and the parallax, $\pi$, we used the default priors in \verb|orbitize!|. Our priors on the total mass and parallax of the system are set to Gaussian priors, with $M_\mathrm{tot}=2.49\pm{}0.27$ \citep{Balmer2022} and $\pi = 6.356 \pm{} 0.047$ \citep{GaiaCollaboration2021}. The MCMC run is set up with 100 walkers and 20 temperatures for a burn-in phase of $50\,000$ steps, followed by an additional $100\,000$ steps to approximate the posterior distributions, keeping only one-tenth of the samples. {The use of multiple temperatures allows the walkers to explore different minimums (at high temperatures), and  to properly sample each available minimum (at lower temperatures).}

\begin{figure*}
    \centering
    \includegraphics[width=\linewidth]{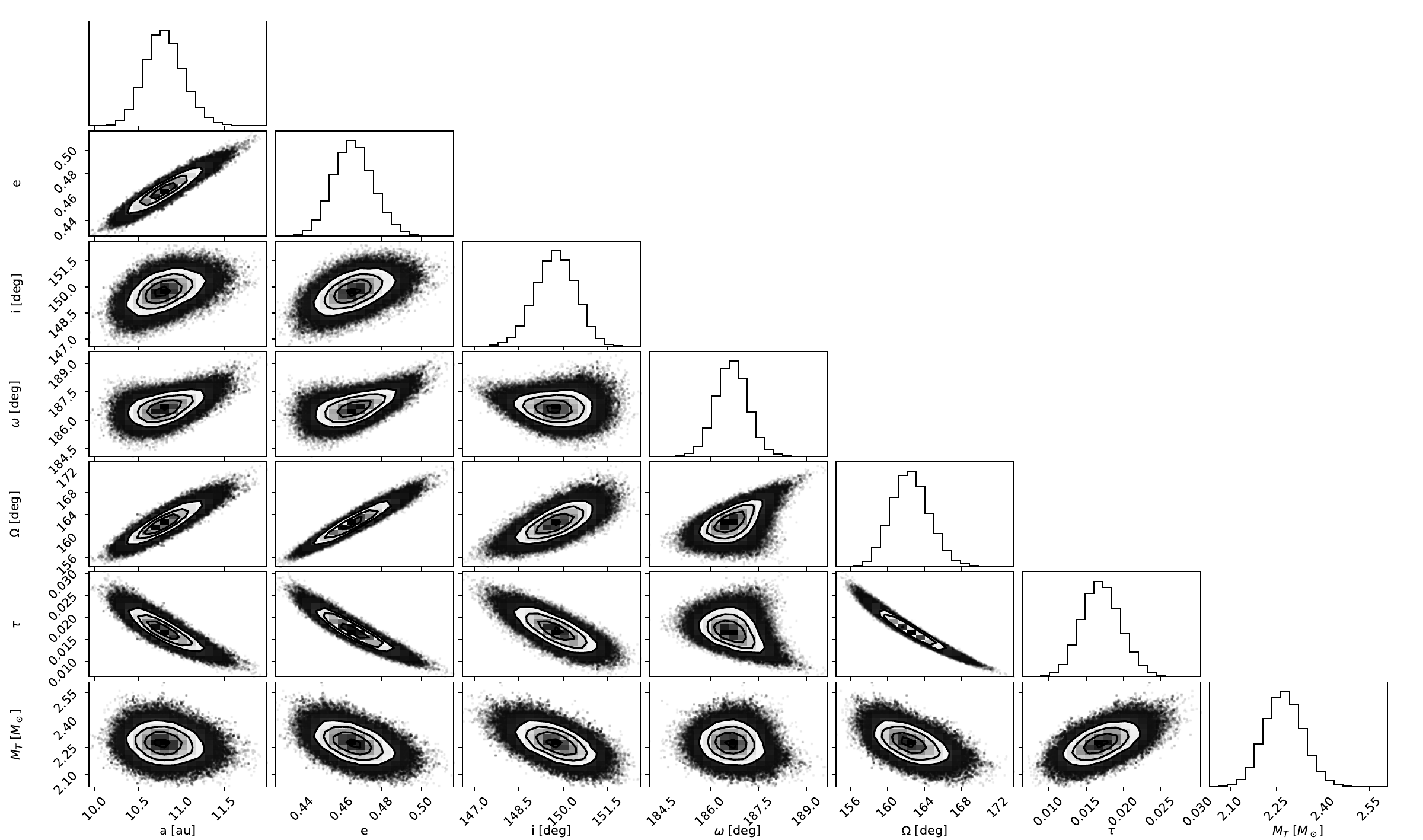}
    \caption{Corner plot for the posterior distributions of the orbital elements of HD 142527 B. The distribution of $\omega_\mathrm{B}$ and $\Omega_\mathrm{B}$ is actually a bimodal distribution, with two peaks spaced 180~deg apart. This is a known degeneracy when radial velocity data are missing. For readability, only one peak, chosen arbitrarily, is shown here.}
    \label{fig:corner}
\end{figure*}

\begin{figure*}
    \centering
    \begin{minipage}[b]{0.49\linewidth}
        \includegraphics[width=\linewidth]{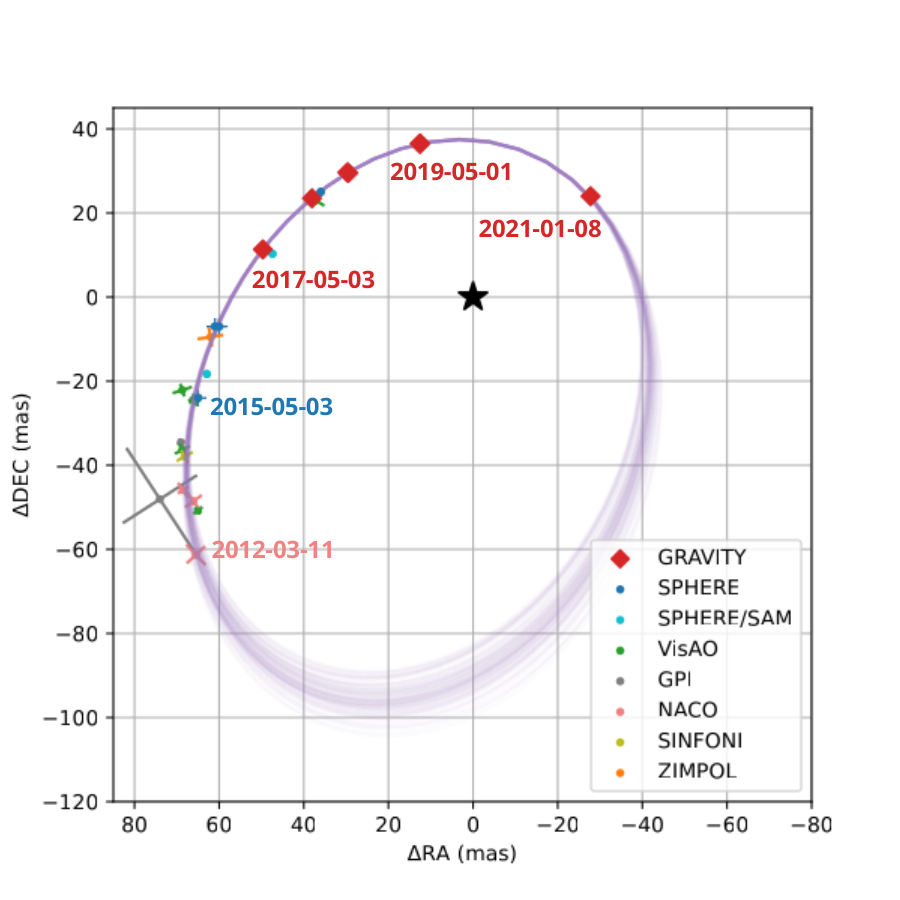}    
    \end{minipage}    
    \begin{minipage}[b]{0.49\linewidth}
        \includegraphics[width=\linewidth, trim=0 0 0 0.8cm, clip=True]{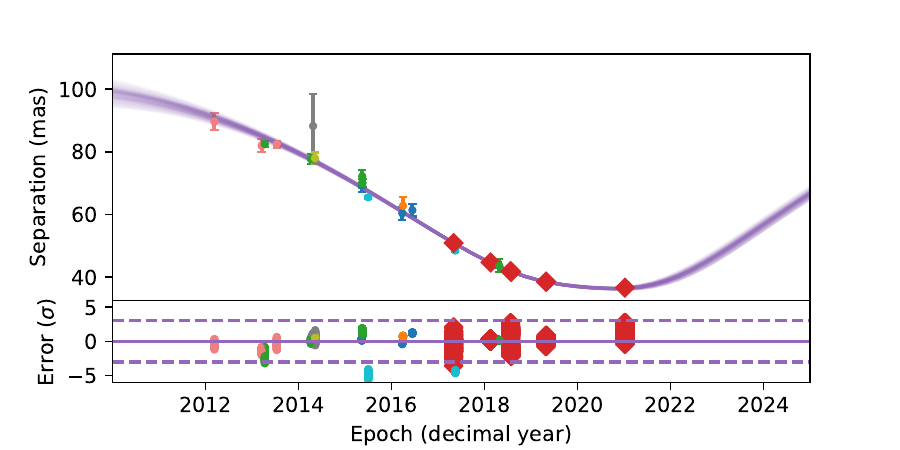}
        \includegraphics[width=\linewidth, trim=0 0 0 0.5cm, clip=True]{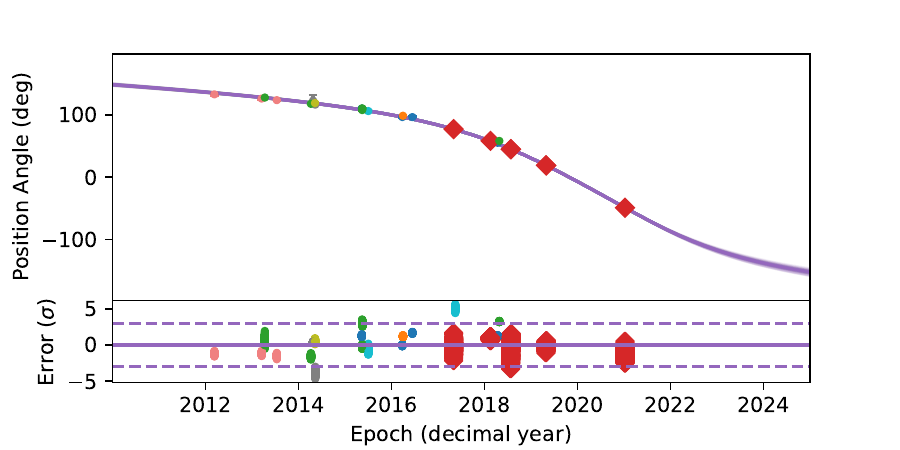}  
    \end{minipage}
    \caption{Illustration of the orbit fit. The left panel shows 100 samples drawn randomly from our posterior distribution, overplotted on the astrometric measurements of Table~\ref{tab:astrometry}. The two panels on the right present the same orbits in separation (upper panel) and position angle (lower panel). For each panel, we show the 100 orbits and the astrometric measurements in the upper graph, and the difference (in units of $\sigma$, the error bars on the measurements) in the lower plot. The two dotted lines show the 3$\sigma$ interval.}
    \label{fig:orbit}
\end{figure*}

Our best estimates for each orbital parameter are reported in Table~\ref{tab:parameters}, and a corner plot is given in Figure~\ref{fig:corner}. In Figure~\ref{fig:orbit} we show a sample of 100 orbits randomly chosen in our posterior distributions, overplotted on the astrometric measurements presented in Table~\ref{tab:astrometry}.

Table~\ref{tab:parameters} and Figure~\ref{fig:orbit} show that the orbit of HD~142527~B is now very well constrained, with a gain of one order of magnitude on the error bars of most orbital parameters compared to previous studies \citep{Lacour2016, Claudi2019, Balmer2022}. The companion is on an eccentric orbit ($e=0.47\pm{0.01}$), with a semi-major axis of $a=10.86\pm{}0.26\,\mathrm{au}$, about 2$\sigma$ from the results of \cite{Balmer2022}. The total mass of the system is $M_\mathrm{tot} = 2.24\pm{}0.06\,M_\odot{}$, resulting in an orbital period of $P=23.95\pm{}0.99\,\mathrm{yr}$. We also find that the perihelion passage occurred on June 02, 2020 (between May 15 and June 21, 2020, at 1$\sigma$). Therefore, our last data point, from January 2021, was obtained after the perihelion passage. This, in combination with the excellent precision of our VLTI/GRAVITY measurements, explains the significant gain in precision on the orbit estimate.

We note that among all the measurements gathered from the literature, the two epochs of SPHERE/SAM data are the only ones that seem to have a systematic and significant disagreement with our orbit fit results. Each of these   astrometric measurements is more than 3$\sigma$ away from the best fit in terms of separation. However, running a second orbit fit with these two data points excluded produced similar results, with all estimates of the orbital parameters within the 1$\sigma$ range of what is obtained with the SPHERE/SAM points. For completeness, we therefore decided to keep these points in our final fit.

\subsection{{Mutual inclination relative to the discs}}

Contradictory values have been reported for the mutual inclination between the orbital plane of the binary and the disc, which is a quantity of particular interest to understand disc--binary interactions. This apparent contradiction seems to stem from a misunderstanding of the multiple conventions used to report disc parameters.

\cite{Czekala2019} reported an inclination relative to the outer disc of $\theta_\mathrm{out} = 35\pm{}5\,\mathrm{deg}$ (or $80\pm{}10\,\mathrm{deg}$\footnote{In absence of radial velocity data, there is a degeneracy in the orbit of the binary, which also leads to a degeneracy in the values of the mutual inclination.}), which they obtained using an inclination $i_\mathrm{out} = 153\,\mathrm{deg}$ and a longitude of the ascending node (LAN) of $\Omega_\mathrm{out} = 160.9\,\mathrm{deg}$ for the disc. \cite{Balmer2022} used the values $i_\mathrm{out} = 38.21~\mathrm{deg}$ and $\Omega_\mathrm{out} = 162.72\,\mathrm{deg}$ from \cite{Bohn2022} to calculate the mutual inclination, and found $\theta_\mathrm{out} = 89.84^{+2.30}_{-1.64}\,\mathrm{deg}$ (or $158.82^{+2.76}_{-2.81}\,\mathrm{deg}$).

The mutual inclination $\theta_\mathrm{out}$ between the binary orbit and the outer disc can be calculated using the   equation \citep{Czekala2019}
\begin{footnotesize}
  \begin{align}
  \cos(\theta_\mathrm{out}) &= \cos(i_\mathrm{out})\cos(i) + \sin(i_\mathrm{out})\sin(i)\cos(\Omega_\mathrm{out} - \Omega), \label{eq:11} 
\end{align}
\end{footnotesize}
\noindent{}where $i_\mathrm{out}$ and $\Omega_\mathrm{out}$ are the inclination and the  LAN  of the outer disc.

It is important to note, though, that in order to be used in Equation~\ref{eq:11},   the parameters of the binary and of the disc must both follow the conventions used in \cite{Czekala2019}. In this convention the inclination is the angle between the axis perpendicular to the plane of sky going towards the observer and the momentum vector of the disc or binary (see their Fig. 1). The LAN is defined as the angle (positive towards the east) between the north and the ascending node, which itself is defined as the point where the disc  crosses the plane of sky receding away from us (i.e.  redshifted).

This convention is widely accepted in the visual binary and exoplanet direct imaging fields, and is the one used in \verb|orbitize!|. However, it seems that the disc parameters reported by \cite{Bohn2022} follow a different convention, in which the inclination is defined as the angle between the plane of sky and the orbital plane.

Two conventions seem to be in use to define disc parameters as utterly different values have been reported for the outer disc of HD~142527. For example, \cite{Casassus2015} use an inclination of $160\,\mathrm{deg}$ in their modelling, similar to the value of $153\,\mathrm{deg}$ reported in \cite{Czekala2019}. These values correspond respectively to an angle between the plane of sky and the orbital plane of 30~deg and 27~deg, more similar to the values reported by \cite{Fukagawa2013} (26.9~deg) and \cite{Bohn2022} ($38.21~\mathrm{deg}$).

Therefore, in order to be used in Equation~\ref{eq:11}, the disc parameters need to be converted to the proper convention. For the outer disc, the parameters inferred by \cite{Bohn2022} from ALMA data correspond to
\begin{align}
  \begin{cases}
    i_\mathrm{out} = 180 - 38.21\pm{}1.38 = 141.79\pm{}1.38\,\mathrm{deg} \\
    \Omega_\mathrm{out} = 162.72\pm{}1.38\,\mathrm{deg}
  \end{cases},
  \label{eq:13}
\end{align}
\noindent{}which corresponds to a disc inclined 38~deg relative to the plane of the sky where the east side is the far side of the disc (see section 4.2 in \citealt{Casassus2015} for a discussion), rotating clockwise on the sky, where the redshifted node is slightly east of south, in agreement with the velocity maps presented in \cite{Bohn2022}.


To determine the values of the mutual inclination of the orbit of HD~142527~B with respect to the two different discs, we use the values of the disc parameters from Equation~\ref{eq:13} together with our posterior distribution of orbits. For each of our $10^6$ sets of posterior orbit ($100\,000$ samples for 100 walkers with a 1/10 thinning), we randomly draw a value for $i_\mathrm{out}, \Omega_\mathrm{out}$ using Gaussian distributions corresponding to Equation~\ref{eq:13}, and we use Equation~\ref{eq:11} to calculate $\theta_\mathrm{out}$. The resulting posterior distributions are bimodal, due to the $180\,\mathrm{deg}$ degeneracy in $\Omega$, and the values are reported in Table~\ref{tab:parameters}.

{We note that the inclination reported by \cite{Bohn2022} for the outer disc is significantly different from previously reported values, which range from 20 to 28~deg \citep{Verhoeff2011, Perez2014, Boehler2017, Garg2021}. This raises the question of whether this is related to the modelling, or if this is a sign of a true physical effect, such as a potentially warped disc. In any case, assuming an inclination of 20~deg (i.e. 160~deg in the reference system of Equation \ref{eq:11}) and a longitude of the ascending node of 160~deg (as in \citealt{Pinte2018}) results in a mutual inclination of 10.5~deg (or 50.5~deg, due to the degeneracy in $\Omega$) relative to the outer disc.}


Contrary to what was presented in \cite{Price2018}, and more recently in \cite{Balmer2022}, who both suggested that the presence HD~142527~B is enough to explain the numerous features observed in the disc (central cavity, spiral arms, asymmetries), our new VLTI/GRAVITY data paint a very different picture. With an eccentric ($e=0.47\pm{}0.01$) but smaller ($a=10.86\pm{}0.26\,\mathrm{au}$) orbit, a maximum separation at aphelion of $15.85\pm{}0.43\,\mathrm{au}$, our new orbit once more raises the question of whether HD~142527~B can really be responsible for the morphology of the disc, and in particular for the large $\simeq{}100\,\mathrm{au}$ observed cavity.

\begin{table}
    \centering
    \begin{tabular}{l l l}
    \hline
    \hline
    Parameter & Value & Unit\\
    \hline
Semi-major axis, $a$ & $10.80\pm0.22$ & au\\
Eccentricity, $e$ & $0.47\pm0.01$ & - \\
Inclination, $i$ & $149.47\pm0.71$ & deg\\
$^1$Argument of periapsis, $\omega$ & $186.45\pm0.48$ & deg \\
$^1$Long. of ascending node, $\Omega$ & $161.51\pm2.01$ & deg \\
Time at periapsis, $\tau$ & $2020.42\pm0.05$ & yr\\
Parallax, $\pi$ & $6.35\pm0.05$ & mas\\
Total mass, $M_\mathrm{tot}$ ($M_\odot{}$) & $2.29\pm0.06$ & $M_\odot{}$\\
\hline
$^2$Period, $P$ & $23.50\pm0.85$ & yr \\
$^2$Distance at apoapsis, $R_\mathrm{apoapsis}$ & $15.85\pm{}0.43$ & au \\
\cdashline{1-3}[0.5pt/3pt]
\multirow{2}{*}{$^{1,2}$Incl. w.r.t. outer disc, $\theta_\mathrm{out}$} & $^{3}$$8.80\pm 1.54$ & \multirow{2}{*}{deg} \\
 &  $^{4}$$2.6-10.5$ & \\
    \hline
    \end{tabular}
    \caption{Orbital elements of HD~142527~B obtained from a fit to the astrometric measurements presented in Table~\ref{tab:astrometry}. \\
    $^1$ The orbital period and mutual inclinations are not direct results of the orbit fit, but are calculated from the posterior distribution. \\
    $^2$ In absence of any radial velocity data, there is a degeneracy  in $\Omega$ and $\omega$, and a second, equally valid, solution can be obtained by adding 180 deg to both $\omega$ and $\Omega$. This also results in a degeneracy in the mutual inclinations, with $\theta_\mathrm{out}=68.45\pm 1.54\,\mathrm{deg}$ or $50.5 - 58.5\,\mathrm{deg}$ being a second set of valid solutions.\\
    $^3$ Using the parameters reported in \cite{Bohn2022} for the outer disc.\\
    $^4$ Using a range of 20 to 28 deg for the inclination of the outer disc, and $\Omega_\mathrm{out} = 160\,\mathrm{deg}$ \citep{Verhoeff2011, Perez2014, Boehler2017, Price2018, Garg2021}.}
    \label{tab:parameters}
\end{table}

\section{Disc--binary interaction}
\label{sec:disc}
\subsection{Hydrodynamical simulation}

To explore whether the new orbit of the companion HD~142527~B is still able to explain the numerous features observed in the disc, we run three 3D gas only hydrodynamical simulations using \textsc{Phantom}, a 3D Smoothed Particle Hydrodynamics (SPH) code \citep{Price2018a}. 

The first simulation recreates one of the simulations in \cite{Price2018} (the R2 orbit). This serves as our control simulation to study the effect the new companion orbit has on the disc. The second simulation uses the same disc parameters as that of \cite{Price2018}, but the updated orbital measurements of the companion as given in Table \ref{tab:parameters}. 
This allows us to easily demonstrate the difference in disc evolution between our new orbit and that used in \citet{Price2018}. {In this set-up, the mutual inclination is $\theta_\mathrm{out} = 50.5\, \mathrm{deg}$, obtained by adding 180 to $\omega$ and $\Omega$ in Table \ref{tab:parameters}}.  The true anomaly is set as $74.31\,\mathrm{deg}$. The third simulation {is identical to the second simulation, except for a reduced} inner boundary of the disc $R_\mathrm{in}$, such that it matches the ratio of $R_\mathrm{in}/R_\mathrm{apoapsis}$ in the first simulation.

All three simulations model a $0.01 M_\odot$ disc with ${R_\mathrm{out} = 350\,\mathrm{au}}$. The surface density profile is given by ${\Sigma \propto R^{-1}}$. The temperature profile is given by ${T \propto R^{-0.3}}$, with ${H/R = 0.11}$ at $R = R_\mathrm{out}$. The inner boundary of the first and second simulations is $R_\mathrm{in} = 50\,\mathrm{au}$, whereas for the third simulation it is $R_\mathrm{in} = 12.67\,\mathrm{au}$. We set a mean Shakura--Sunyaev disc viscosity $\alpha_\mathrm{SS} \approx 0.005$. This corresponds to an artificial viscosity $\alpha_\mathrm{AV} = 0.22$ for the first and second simulation, and $\alpha_\mathrm{AV} = 0.18$ for the third simulation. The artificial viscosity coefficient $\beta_{\text{AV}}$ is set to 2 (see \citealt{Price2018a,2015Nealon}).

Sink particles are used to model the binary \citep{1995Bate}. The mass and orbital parameters of the binary for the first simulation are identical to the R2 orbit in \cite{Price2018}, whereas for the second and third simulations the mass of the primary and secondary is $2.03 M_\odot$ and $0.26 M_\odot$, respectively, consistent with \cite{Claudi2019} and the total mass in Table \ref{tab:parameters}. The accretion radius for both sinks is set to $1\textsc{au}$, as in \cite{Price2018}. The simulation recreating \cite{Price2018} is shown at 20 orbital periods (3272 yr) corresponding to their Figure 2. We present all of our results at the same time. For the updated companion orbit, this corresponds to 141 orbital periods.

\begin{figure*}
    \centering
    \includegraphics[width=\linewidth]{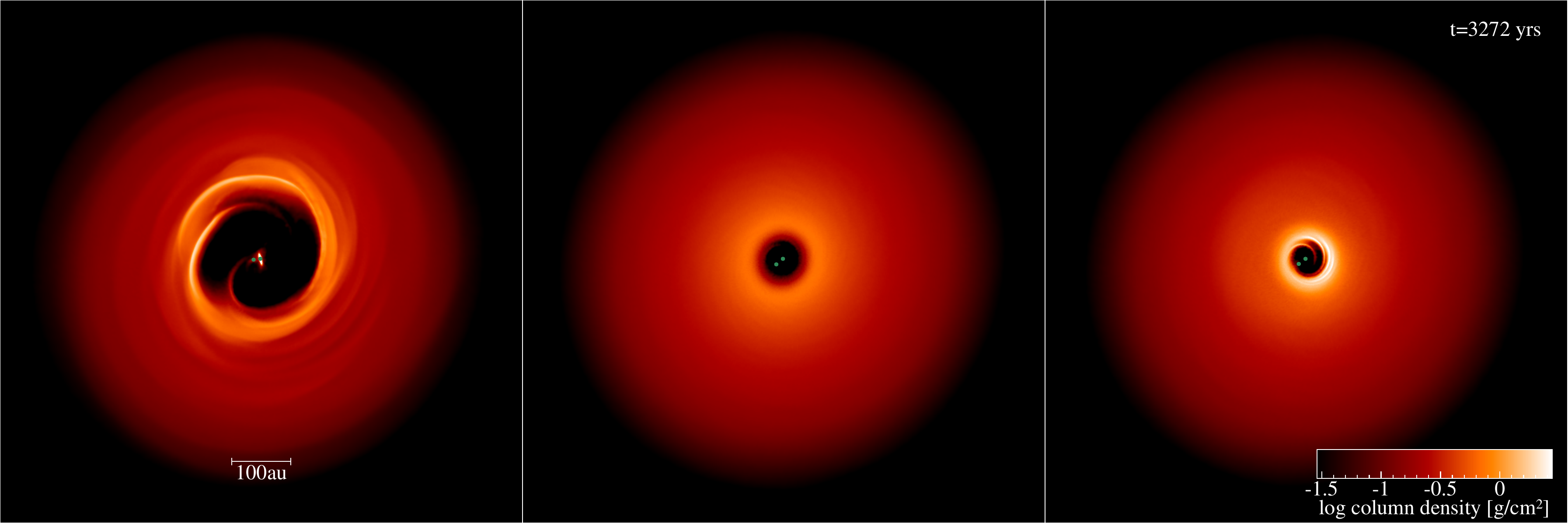}
    \caption{Surface density rendered plots of the HD~142527 protoplanetary disc. Each panel shows a different simulation at the same moment in time, but with different disc and binary parameters. The left panel reproduces the results of the R2 orbit in \cite{Price2018} where the companion was shown to be responsible for the observed disc features: a large cavity, prominent spiral arms, and a misaligned inner disc. The middle panel uses the same disc parameters as the left panel, but uses the updated binary parameters in Table \ref{tab:parameters}. The right panel is set up with the {second solution of the} updated binary parameters, and a smaller inner disc boundary to match the ratio of $R_\mathrm{in}/R_\mathrm{apoapsis}$ of the simulation in the left panel. In the middle and right panels we find that, due to the close configuration of the updated binary orbit, the companion cannot be solely responsible for the observed disc features.}
    \label{fig:discFeatures}
\end{figure*}

Figure \ref{fig:discFeatures} shows all three of the   simulations at the same moment in time. The left panel shows the simulation using the R2 orbit set-up from \cite{Price2018}, which was able to explain the observed features of the disc. In this orbit set-up, the companion is responsible for the large cavity (${\sim}100\,\mathrm{au}$), the large spiral arms, and the misaligned inner disc casting a shadow. The middle panel shows the simulation where the only change is the companion's orbit. As we have reproduced the results of \cite{Price2018}, we can attribute all differences in the disc structure to the change in the orbital measurements. The new orbit is unable to match any of the disc features mentioned above. The large separation between the inner edge of the disc and the companion results in little impact on the disc structure, resulting in a much smaller cavity \citep{Bohn2022}. Finally, the right panel shows the simulation where the inner boundary of the disc is closer in, such that the ratio $R_\mathrm{in}/R_\mathrm{apoapsis}$ is equivalent to the original simulation from \cite{Price2018} (left panel). While there are improvements, the companion is still unable to carve open a large enough cavity. Furthermore, the lack of a pressure maximum to trap dust at ${\sim}100\,\mathrm{au}$ makes it difficult to explain the observed dust structures \citep{Casassus2015}.

The inability of the companion to clear a large enough cavity is unsurprising. From theoretical predictions \citep{1994Artymowicz}, a circular binary with $\mu =  0.1$ as in this system,  is expected to result in a disc inner edge of $r=1.74a$, where $\mu = M_\mathrm{companion}/M_\mathrm{tot}$ is the mass fraction of the perturber. Although eccentric companions result in larger cavities, there is no prediction where the companion is able to carve open a big enough cavity to match observations. Figure 4 in \cite{1994Artymowicz} shows that for a binary with $\mu = 0.1$, an eccentricity of 0.47, and an assumed disc viscosity of $5 \times 10^{-3}$, the disc inner edge is expected to be between $r = 2.7a$ and $3a$. For a semi-major axis of $a=10.8\,\mathrm{au}$, this corresponds to a disc inner edge at $r \approx 30\,\mathrm{au}$, which agrees well with the simulations in the middle and right panels of Figure \ref{fig:discFeatures}. Additionally, the disc is inclined relative to the orbital plane of the binary, which decreases the truncation radius \citep{2015Lubow,2018Facchini}. The observed cavity size of ${\sim} 100\,\mathrm{au}$ far exceeds even the most generous predictions. Hence, the simulations here show that with the new orbital parameters, the companion cannot be solely responsible for the observed disc features.

\begin{figure*}
    \centering
    \includegraphics[width=0.66\linewidth]{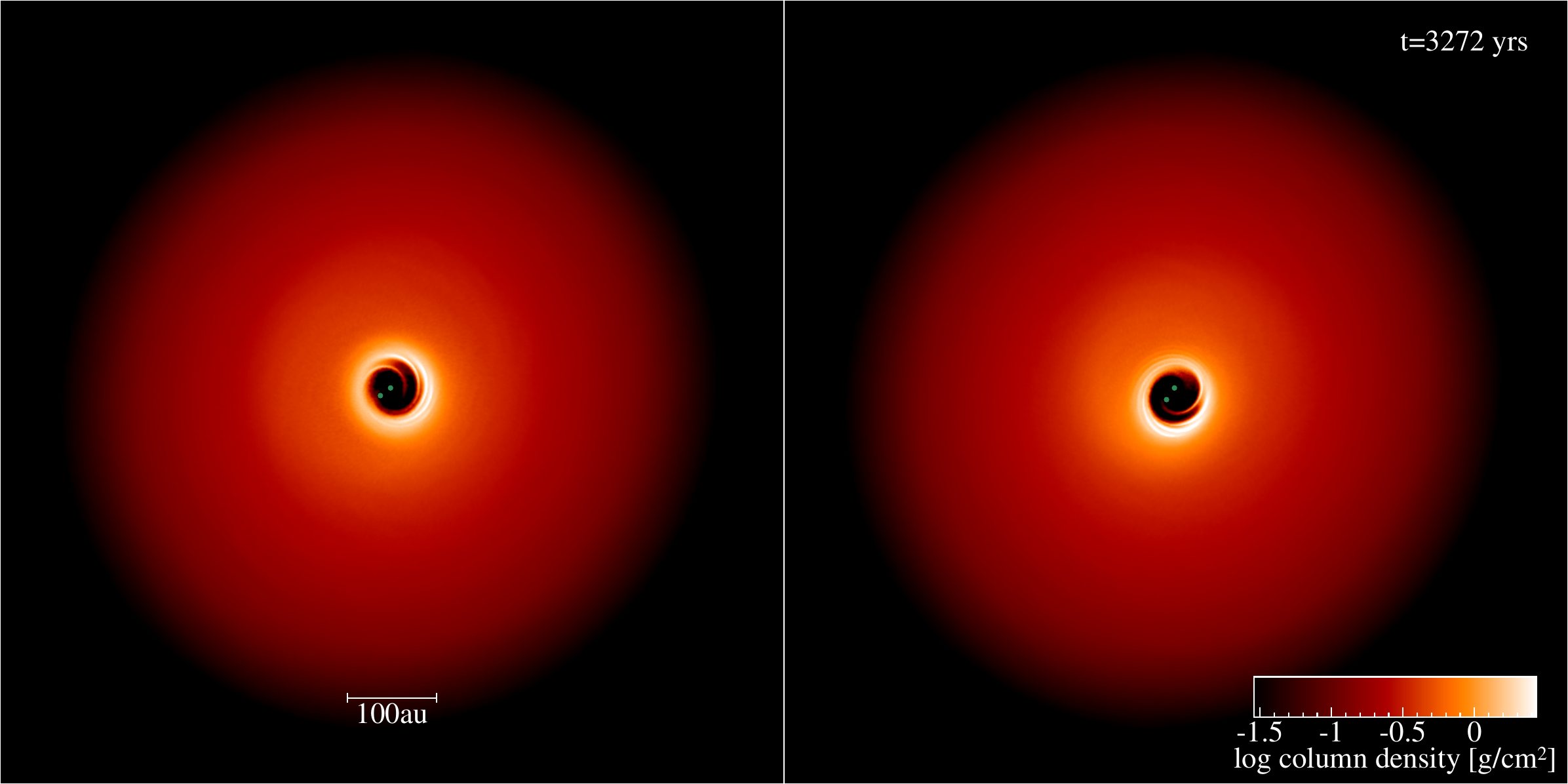}
    \caption{Surface density rendered plots of the HD~142527 protoplanetary disc. Each panel shows a simulation at the same moment in time, but with different disc and binary parameters. The {right} panel is set up with the updated binary parameters as listed in Table \ref{tab:parameters}. The {left} panel uses the second solution of the values in \ref{tab:parameters} obtained by adding $180\,\mathrm{deg}$ to both $\Omega$ and $\omega$. We find that neither solution reproduces the observed disc features.}
    \label{fig:altSolution}
\end{figure*}

For completeness, we also simulated the solution of the binary orbit using  the values of $\Omega$ and $\omega$ listed in Table \ref{tab:parameters} {resulting in  $\theta_\mathrm{out} = 10.5\, \mathrm{deg}$}.  Figure \ref{fig:altSolution} shows that neither solution is able to reproduce the disc features. {Since both solutions produce similar results, we only use the simulation with the second solution (left panel) for the comparisons with the observations below.}

\subsection{Radiative transfer}

\begin{figure*}
    \centering
    \includegraphics[width=0.9\linewidth]{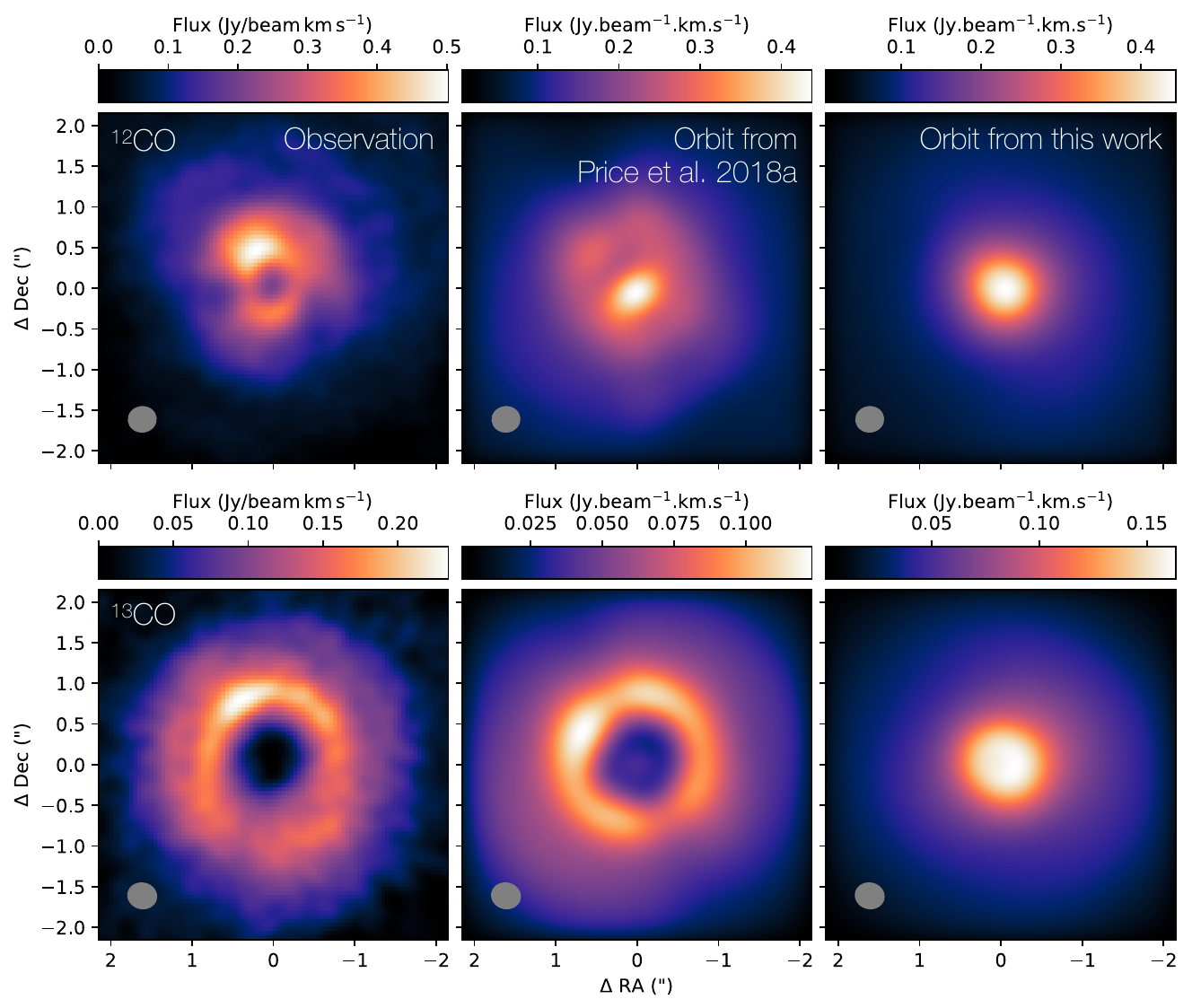}
    \caption{Comparison between observed and simulated $^{12}$CO and $^{13}$CO maps for the HD142527 disc. The top and bottom panels show the $^{12}$CO and $^{13}$CO $\text{J} = 2-1$ emission, respectively. The left panels are the observations \citep{Garg2021}. The middle and right panels are synthetic emissions predicted from the simulations. The middle panels are from the simulation shown in the left panel of Figure \ref{fig:discFeatures} which uses the disc and binary set-up as the R2 orbit in \cite{Price2018}. The right panels are from the simulations shown in the right panel of Figure \ref{fig:discFeatures}, which uses the updated binary parameters in Table \ref{tab:parameters}. In general, the new binary orbit results in a lack of substructure in either molecule. Additionally, the lack of a cavity in $^{13}$CO further highlights that the companion alone cannot reproduce the disc features.}
    \label{fig:CO}
\end{figure*}

For  comparison with observations of the $^{12}$CO and $^{13}$CO ${\text{J} = 2-1}$ line emission, we use the radiative transfer code \textsc{mcfost} \citep{2006Pinte,2009Pinte}. As it uses a Voronoi tesselation where each cell in \textsc{mcfost} corresponds to an individual SPH particle, \textsc{mcfost} is well suited for post-processing SPH simulations.

A subset of the simulations above are post-processed, in particular the simulations using the binary orbit from \cite{Price2018} and the updated binary orbit from this work. The sink particles are used to irradiate the disc using  stellar spectral models based on their mass, and assuming a 3Myr isochrone. For the former simulation, this corresponds to the luminosity and effective temperature of ${L = 2.69 L_\odot}$ and ${T_\text{eff} = 4813K}$, respectively, for the primary, and ${L = 0.23} L_\odot$ and ${T_\text{eff} = 3607 K}$, respectively, for the secondary. For the latter simulation, the luminosity and effective temperature is ${L = 3.89 L_\odot}$ and ${T_\text{eff} = 4991 K}$, respectively, for the primary, and ${L = 0.14 L_\odot}$ and ${T_\text{eff} = 3315 K}$, respectively, for the secondary.

We use $10^8$ photon packets to calculate the temperatures in the disc. As these are gas-only simulations, the dust is assumed to be  $1\%$ of the gas mass, and perfectly coupled to the gas. The dust sizes vary between 0.3 and 1000$\mu m$, and are distributed across 100 grain sizes with a power-law exponent of $-3.5$. We further assume that the dust particles are all spherical and homogeneous, and made up of astronomical silicates. The dust properties are computed using Mie theory. To predict the line emission of $^{12}$CO and $^{13}$CO, we assume constant abundances of $5 \times 10^{-5}$ and $7 \times 10^{-7}$, respectively.

Figure \ref{fig:CO} compares the observations (left panels) with the predictions from simulations (middle and right panels). The top and bottom panels show the $^{12}$CO and $^{13}$CO ${\text{J} = 2-1}$ line emission, respectively. In general, the new binary orbit results in a lack of substructure with either molecule as seen by the right panels. In $^{13}$CO, the new orbit is unable to deplete the gas in the inner regions, highlighted by the lack of any cavity compared to either the observation (bottom left panel) or the simulation from \cite{Price2018} (bottom middle panel). Although the $^{12}$CO emission is lacking any substructure with the new binary orbit, it is more similar   to the observation (top left panel) and the simulation \cite{Price2018} (top middle panel); all have an optically thick inner region. While there is a lot of gas in the cavity, this could clear out if the simulation evolved for longer. Additionally, the beam size washed out the cavity seen in the surface density.

\section{Conclusion}
\label{sec:conclusion}

Based on a VLTI/GRAVITY observations, we presented a set of new astrometric measurements of HD~142527~B relative to HD~142527~B obtained at seven epochs ranging from 2017 to early 2021. Combined with all the measurements available in the literature, this constitutes an astrometric monitoring of $\sim{}9\,\mathrm{yr}$, which covers almost half of the orbital period of the binary.

From these measurements, we derived excellent estimates of the orbital parameters for the binary. We find that HD~142527~B is on a moderately eccentric orbit ($e=0.47\pm{}0.01$), with a semi-major axis of $10.80\pm{}0.22\,\mathrm{au}$. The orbital period is therefore $P=23.50\pm{}0.85\,\mathrm{yr}$, and we estimate that the low-mass companion passed its perihelion on June 02, 2020. We also report a total mass for the system of $M_\mathrm{tot}=2.29\pm{}0.06M_\odot{}$.

Considering our constraints on the orbital parameters of HD~142527~B, we calculate that its maximum separation with respect to the central star HD~142527~A is $R_\mathrm{apoapsis}=15.85\pm{}0.43\,\mathrm{au}$. Contrary to what has been previously suggested \citep{Price2018} based on different orbital estimates, this makes it highly unlikely that this low-mass companion can be solely responsible for the large $\sim{}100\,\mathrm{au}$ cavity observed in the  near- to mid-infrared \citep{Fukagawa2006, Verhoeff2011} and more recently with ALMA \citep{Casassus2013}.

To confirm this, we ran a set of three gas-only 3D hydrodynamical simulations of the disc around HD~142527 AB. We used the first simulation set-up as a control to demonstrate that we were able to reproduce the results obtained by \cite{Price2018} under the same set of assumption regarding the orbital parameters of the binary. In the second simulation, we switched the orbit of the binary for our new estimate, based on the newly available VLTI/GRAVITY observations. We were unable to reproduce most of the features observed in the disc with this set-up: the low-mass companion HD~142527~B could only carve a $\sim{}30\,\mathrm{au}$ cavity, and did not 
interact with the disc. In the third set-up, we reduced the inner radius of the disc, allowing the companion to have a greater impact on the disc structure. However, the companion alone was {still} unable to explain the disc features.

A larger theoretical follow-up will be necessary to determine what else is required to match this new orbit of HD~142527 B with the observed disc structures. Additional observations could also prove useful as they could potentially reveal the presence of an additional companion in the system, which would help explain the disc structures and/or the presence of additional material extending further in the disc around HD~142527.


\begin{acknowledgements}
The authors thank Grant Kennedy for a helpful discussion on the calculation of mutual inclinations and the conventions used for expressing disc parameters. 
This work is based on observations collected at the European Southern Observatory under ESO programmes 098.D-0488, 60.A-9800, 60.A-9132 and 1104.C-0651.
Most simulations in this work were performed using the DiRAC Data Intensive service at Leicester, operated by the University of Leicester IT Services, which forms part of the STFC DiRAC HPC Facility. The equipment was funded by BEIS capital funding via STFC capital grants ST/K000373/1 and ST/R002363/1 and STFC DiRAC Operations grant ST/R001014/1. DiRAC is part of the National e-Infrastructure.
S. R acknowledges support from the Science \& Technology Facilities Council (STFC) through Consolidated Grant ST/W000857/1.
R. N. acknowledges support from UKRI/EPSRC through a Stephen Hawking Fellowship (EP/T017287/1).
S. L. acknowledges the support of the French Agence Nationale de la Recherche (ANR), under grant ANR-21-CE31-0017 (project ExoVLTI). 
D. P. acknowledges Australian Research Council funding via DP180104235.
\end{acknowledgements}

\bibliographystyle{aa}
\bibliography{biblio}

\begin{thebibliography}{40}
\expandafter\ifx\csname natexlab\endcsname\relax\def\natexlab#1{#1}\fi

\bibitem[{{Artymowicz} \& {Lubow}(1994)}]{1994Artymowicz}
{Artymowicz}, P. \& {Lubow}, S.~H. 1994, \apj, 421, 651

\bibitem[{Avenhaus {et~al.}(2017)Avenhaus, Quanz, Schmid, Dominik, Stolker,
  Ginski, de~Boer, Szul{\'a}gyi, Garufi, Zurlo, Hagelberg, Benisty, Henning,
  M{\'e}nard, Meyer, Baruffolo, Bazzon, Beuzit, Costille, Dohlen, Girard,
  Gisler, Kasper, Mouillet, Pragt, Roelfsema, Salasnich, \&
  Sauvage}]{Avenhaus2017}
Avenhaus, H., Quanz, S.~P., Schmid, H.~M., {et~al.} 2017, AJ, 154, 33

\bibitem[{Avenhaus {et~al.}(2014)Avenhaus, Quanz, Schmid, Meyer, Garufi, Wolf,
  \& Dominik}]{Avenhaus2014}
Avenhaus, H., Quanz, S.~P., Schmid, H.~M., {et~al.} 2014, ApJ, 781, 87

\bibitem[{Balmer {et~al.}(2022)Balmer, Follette, Close, Males, Rosa, Redai,
  Watson, Weinberger, Morzinski, Morales, {Ward-Duong}, \& Pueyo}]{Balmer2022}
Balmer, W.~O., Follette, K.~B., Close, L.~M., {et~al.} 2022, AJ, 164, 29

\bibitem[{{Bate} {et~al.}(1995){Bate}, {Bonnell}, \& {Price}}]{1995Bate}
{Bate}, M.~R., {Bonnell}, I.~A., \& {Price}, N.~M. 1995, \mnras, 277, 362

\bibitem[{Biller {et~al.}(2012)Biller, Lacour, Juh{\'a}sz, Benisty, Chauvin,
  Olofsson, Pott, M{\"u}ller, {Sicilia-Aguilar}, Bonnefoy, Tuthill, Thebault,
  Henning, \& Crida}]{Biller2012}
Biller, B., Lacour, S., Juh{\'a}sz, A., {et~al.} 2012, ApJL, 753, L38

\bibitem[{Blunt {et~al.}(2020)Blunt, Wang, Angelo, Ngo, Cody, Rosa, Graham,
  Hirsch, Nagpal, Nielsen, Pearce, Rice, \& Tejada}]{Blunt2020}
Blunt, S., Wang, J.~J., Angelo, I., {et~al.} 2020, AJ, 159, 89

\bibitem[{Boehler {et~al.}(2017)Boehler, Weaver, Isella, Ricci, Grady,
  Carpenter, \& Perez}]{Boehler2017}
Boehler, Y., Weaver, E., Isella, A., {et~al.} 2017, ApJ, 840, 60

\bibitem[{Bohn {et~al.}(2022)Bohn, Benisty, Perraut, van~der Marel, W{\"o}lfer,
  van Dishoeck, Facchini, Manara, Teague, Francis, Berger, {Garcia-Lopez},
  Ginski, Henning, Kenworthy, Kraus, M{\'e}nard, M{\'e}rand, \&
  P{\'e}rez}]{Bohn2022}
Bohn, A.~J., Benisty, M., Perraut, K., {et~al.} 2022, A\&A, 658, A183

\bibitem[{Canovas {et~al.}(2013)Canovas, M{\'e}nard, Hales, Jord{\'a}n,
  Schreiber, Casassus, Gledhill, \& Pinte}]{Canovas2013}
Canovas, H., M{\'e}nard, F., Hales, A., {et~al.} 2013, A\&A, 556, A123

\bibitem[{Casassus {et~al.}(2013)Casassus, Hales, de~Gregorio, Dent, Belloche,
  G{\"u}sten, M{\'e}nard, Hughes, Wilner, \& Salinas}]{Casassus2013}
Casassus, S., Hales, A., de~Gregorio, I., {et~al.} 2013, A\&A, 553, A64

\bibitem[{Casassus {et~al.}(2012)Casassus, M, Jord{\'a}n, M{\'e}nard, Cuadra,
  Schreiber, Hales, \& Ercolano}]{Casassus2012}
Casassus, S., M, S.~P., Jord{\'a}n, A., {et~al.} 2012, ApJL, 754, L31

\bibitem[{Casassus {et~al.}(2015)Casassus, Marino, P{\'e}rez, Roman, Dunhill,
  Armitage, Cuadra, Wootten, van~der Plas, Cieza, Moral, Christiaens, \&
  Montesinos}]{Casassus2015}
Casassus, S., Marino, S., P{\'e}rez, S., {et~al.} 2015, ApJ, 811, 92

\bibitem[{Christiaens {et~al.}(2014)Christiaens, Casassus, Perez, van~der Plas,
  \& M{\'e}nard}]{Christiaens2014}
Christiaens, V., Casassus, S., Perez, S., van~der Plas, G., \& M{\'e}nard, F.
  2014, ApJL, 785, L12

\bibitem[{Claudi {et~al.}(2019)Claudi, Maire, Mesa, Cheetham, Fontanive,
  Gratton, Zurlo, Avenhaus, Bhowmik, Biller, Boccaletti, Bonavita, Bonnefoy,
  Cascone, Chauvin, Delboulb{\'e}, Desidera, D'Orazi, Feautrier, Feldt, Dotti,
  Girard, Giro, Janson, Hagelberg, Keppler, Kopytova, Lacour, Lagrange,
  Langlois, Lannier, Coroller, Menard, Messina, Meyer, Millward, Olofsson,
  Pavlov, Peretti, Perrot, Pinte, Pragt, Ramos, Rochat, Rodet, Roelfsema,
  Rouan, Salter, Schmidt, Sissa, Thebault, Udry, \& Vigan}]{Claudi2019}
Claudi, R., Maire, A.-L., Mesa, D., {et~al.} 2019, A\&A, 622, A96

\bibitem[{Czekala {et~al.}(2019)Czekala, Chiang, Andrews, Jensen, Torres,
  Wilner, Stassun, \& Macintosh}]{Czekala2019}
Czekala, I., Chiang, E., Andrews, S.~M., {et~al.} 2019, ApJ, 883, 22

\bibitem[{{Facchini} {et~al.}(2018){Facchini}, {Juh{\'a}sz}, \&
  {Lodato}}]{2018Facchini}
{Facchini}, S., {Juh{\'a}sz}, A., \& {Lodato}, G. 2018, \mnras, 473, 4459

\bibitem[{{Foreman-Mackey} {et~al.}(2013){Foreman-Mackey}, Hogg, Lang, \&
  Goodman}]{Foreman-Mackey2013}
{Foreman-Mackey}, D., Hogg, D.~W., Lang, D., \& Goodman, J. 2013, PASP, 125,
  306

\bibitem[{Fukagawa {et~al.}(2006)Fukagawa, Tamura, Itoh, Kudo, Imaeda, Oasa,
  Hayashi, \& Hayashi}]{Fukagawa2006}
Fukagawa, M., Tamura, M., Itoh, Y., {et~al.} 2006, ApJ, 636, L153

\bibitem[{Fukagawa {et~al.}(2013)Fukagawa, Tsukagoshi, Momose, Saigo, Ohashi,
  Kitamura, Inutsuka, Muto, Nomura, Takeuchi, Kobayashi, Hanawa, Akiyama,
  Honda, Fujiwara, Kataoka, Takahashi, \& Shibai}]{Fukagawa2013}
Fukagawa, M., Tsukagoshi, T., Momose, M., {et~al.} 2013, Publications of the
  Astronomical Society of Japan, 65

\bibitem[{{GaiaCollaboration} {et~al.}(2021){GaiaCollaboration}, Brown,
  Vallenari, Prusti, de~Bruijne, Babusiaux, Biermann, Creevey, Evans, Eyer,
  Hutton, Jansen, Jordi, Klioner, Lammers, Lindegren, Luri, Mignard, Panem,
  Pourbaix, Randich, Sartoretti, Soubiran, Walton, Arenou, {Bailer-Jones},
  Bastian, Cropper, Drimmel, Katz, Lattanzi, van Leeuwen, Bakker, Cacciari,
  Casta{\~n}eda, Angeli, Ducourant, Fabricius, Fouesneau, Fr{\'e}mat, Guerra,
  Guerrier, Guiraud, Piccolo, Masana, Messineo, Mowlavi, Nicolas, Nienartowicz,
  Pailler, Panuzzo, Riclet, Roux, Seabroke, Sordo, Tanga, Th{\'e}venin,
  {Gracia-Abril}, Portell, Teyssier, Altmann, Andrae, {Bellas-Velidis}, Benson,
  Berthier, Blomme, Brugaletta, Burgess, Busso, Carry, Cellino, Cheek,
  Clementini, Damerdji, Davidson, Delchambre, Dell'Oro,
  {Fern{\'a}ndez-Hern{\'a}ndez}, Galluccio, {Garc{\'i}a-Lario},
  {Garcia-Reinaldos}, {Gonz{\'a}lez-N{\'u}{\~n}ez}, Gosset, Haigron, Halbwachs,
  Hambly, Harrison, Hatzidimitriou, Heiter, Hern{\'a}ndez, Hestroffer, Hodgkin,
  Holl, Jan{\ss}en, de~Fombelle, Jordan, {Krone-Martins}, Lanzafame,
  L{\"o}ffler, Lorca, Manteiga, Marchal, Marrese, Moitinho, Mora, Muinonen,
  Osborne, Pancino, Pauwels, Petit, {Recio-Blanco}, Richards, Riello,
  Rimoldini, Robin, Roegiers, Rybizki, Sarro, Siopis, Smith, Sozzetti, Ulla,
  Utrilla, van Leeuwen, van Reeven, Abbas, Aramburu, Accart, Aerts, Aguado,
  Ajaj, Altavilla, {\'A}lvarez, {Cid-Fuentes}, Alves, Anderson, Varela, Antoja,
  Audard, Baines, Baker, {Balaguer-N{\'u}{\~n}ez}, Balbinot, Balog, Barache,
  Barbato, Barros, Barstow, Bartolom{\'e}, Bassilana, Bauchet,
  {Baudesson-Stella}, Becciani, Bellazzini, Bernet, Bertone, Bianchi,
  {Blanco-Cuaresma}, Boch, Bombrun, Bossini, Bouquillon, Bragaglia, Bramante,
  Breedt, Bressan, Brouillet, Bucciarelli, Burlacu, Busonero, Butkevich, Buzzi,
  Caffau, Cancelliere, C{\'a}novas, {Cantat-Gaudin}, Carballo, Carlucci,
  Carnerero, Carrasco, Casamiquela, Castellani, {Castro-Ginard}, Sampol,
  Chaoul, Charlot, Chemin, Chiavassa, Cioni, Comoretto, Cooper, Cornez, Cowell,
  Crifo, Crosta, Crowley, Dafonte, Dapergolas, David, David, de~Laverny, Luise,
  March, Ridder, de~Souza, de~Teodoro, de~Torres, del Peloso, del Pozo, Delbo,
  Delgado, Delgado, Delisle, Matteo, Diakite, Diener, Distefano, Dolding,
  Eappachen, Edvardsson, Enke, Esquej, Fabre, Fabrizio, Faigler, Fedorets,
  Fernique, Fienga, Figueras, Fouron, Fragkoudi, Fraile, Franke, Gai, Garabato,
  {Garcia-Gutierrez}, {Garc{\'i}a-Torres}, Garofalo, Gavras, Gerlach, Geyer,
  Giacobbe, Gilmore, Girona, Giuffrida, Gomel, Gomez, {Gonzalez-Santamaria},
  {Gonz{\'a}lez-Vidal}, Granvik, {Guti{\'e}rrez-S{\'a}nchez}, Guy, Hauser,
  Haywood, Helmi, Hidalgo, Hilger, H{\l}adczuk, Hobbs, Holland, Huckle,
  Jasniewicz, Jonker, Campillo, Julbe, Karbevska, Kervella, Khanna, Kochoska,
  Kontizas, Kordopatis, Korn, {Kostrzewa-Rutkowska}, Kruszy{\'n}ska, Lambert,
  Lanza, Lasne, Campion, Fustec, Lebreton, Lebzelter, Leccia, Leclerc,
  {Lecoeur-Taibi}, Liao, Licata, Lindstr{\o}m, Lister, Livanou, Lobel, Pardo,
  Managau, Mann, Marchant, Marconi, Santos, Marinoni, Marocco, Marshall, Polo,
  {Mart{\'i}n-Fleitas}, Masip, Massari, {Mastrobuono-Battisti}, Mazeh,
  McMillan, Messina, Michalik, Millar, Mints, Molina, Molinaro, Moln{\'a}r,
  Montegriffo, Mor, Morbidelli, Morel, Morris, Mulone, Munoz, Muraveva, Murphy,
  Musella, Noval, Ord{\'e}novic, Orr{\`u}, Osinde, Pagani, Pagano, Palaversa,
  Palicio, Panahi, Pawlak, Esteller, Penttil{\"a}, Piersimoni, Pineau, Plachy,
  Plum, Poggio, Poretti, Poujoulet, Pr{\v s}a, Pulone, Racero, Ragaini, Rainer,
  Raiteri, Rambaux, Ramos, {Ramos-Lerate}, Fiorentin, Regibo, Reyl{\'e},
  Ripepi, Riva, Rixon, Robichon, Robin, Roelens, Rohrbasser,
  {Romero-G{\'o}mez}, Rowell, Royer, Rybicki, Sadowski, Sell{\'e}s, Sahlmann,
  Salgado, Salguero, Samaras, Gimenez, Sanna, Santove{\~n}a, Sarasso,
  Schultheis, Sciacca, Segol, Segovia, S{\'e}gransan, Semeux, Shahaf, Siddiqui,
  Siebert, Siltala, Slezak, Smart, Solano, Solitro, Souami, Souchay, Spagna,
  Spoto, Steele, Steidelm{\"u}ller, Stephenson, S{\"u}veges, Szabados,
  {Szegedi-Elek}, Taris, Tauran, Taylor, Teixeira, Thuillot, Tonello, Torra,
  Torra, Turon, Unger, Vaillant, van Dillen, Vanel, Vecchiato, Viala, Vicente,
  Voutsinas, Weiler, Wevers, Wyrzykowski, Yoldas, Yvard, Zhao, Zorec, Zucker,
  Zurbach, \& Zwitter}]{GaiaCollaboration2021}
{GaiaCollaboration}, Brown, A. G.~A., Vallenari, A., {et~al.} 2021, A\&A, 649,
  A1

\bibitem[{{Garg} {et~al.}(2021){Garg}, {Pinte}, {Christiaens}, {Price},
  {Lazendic}, {Boehler}, {Casassus}, {Marino}, {Perez}, \& {Zuleta}}]{Garg2021}
{Garg}, H., {Pinte}, C., {Christiaens}, V., {et~al.} 2021, \mnras, 504, 782

\bibitem[{{GravityCollaboration} {et~al.}(2019){GravityCollaboration}, Lacour,
  Nowak, Wang, Pfuhl, Eisenhauer, Abuter, Amorim, Anugu, Benisty, Berger,
  Beust, Blind, Bonnefoy, Bonnet, Bourget, Brandner, Buron, Collin, Charnay,
  Chapron, Cl{\'e}net, du~Foresto, {de Zeeuw}, Deen, Dembet, Dexter, Duvert,
  Eckart, Schreiber, F{\'e}dou, Garcia, Lopez, Gao, Gendron, Genzel, Gillessen,
  Gordo, Greenbaum, Habibi, Haubois, Hau{\ss}mann, Henning, Hippler, Horrobin,
  Hubert, Rosales, Jocou, Kendrew, Kervella, Kolb, Lagrange, Lapeyr{\`e}re,
  Bouquin, L{\'e}na, Lippa, Lenzen, Maire, Molli{\`e}re, Ott, Paumard, Perraut,
  Perrin, Pueyo, Rabien, Ramirez, Rau, {Rodriguez-Coira}, Rousset,
  {Sanchez-Bermudez}, Scheithauer, Schuhler, Straub, Straubmeier, Sturm,
  Tacconi, Vincent, {van Dishoeck}, {von Fellenberg}, Wank, Waisberg, Widmann,
  Wieprecht, Wiest, Wiezorrek, Woillez, Yazici, Ziegler, \&
  Zins}]{GravityCollaboration2019}
{GravityCollaboration}, Lacour, S., Nowak, M., {et~al.} 2019, A\&A, 623, L11

\bibitem[{{GravityCollaboration} {et~al.}(2020){GravityCollaboration}, Nowak,
  Lacour, Molli{\`e}re, Wang, Charnay, van Dishoeck, Abuter, Amorim, Berger,
  Beust, Bonnefoy, Bonnet, Brandner, Buron, Cantalloube, Collin, Chapron,
  Cl{\'e}net, du~Foresto, de~Zeeuw, Dembet, Dexter, Duvert, Eckart, Eisenhauer,
  Schreiber, F{\'e}dou, Lopez, Gao, Gendron, Genzel, Gillessen, Hau{\ss}mann,
  Henning, Hippler, Hubert, Jocou, Kervella, Lagrange, Lapeyr{\`e}re, Bouquin,
  L{\'e}na, Maire, Ott, Paumard, Paladini, Perraut, Perrin, Pueyo, Pfuhl,
  Rabien, Rau, {Rodr{\'i}guez-Coira}, Rousset, Scheithauer, Shangguan, Straub,
  Straubmeier, Sturm, Tacconi, Vincent, Widmann, Wieprecht, Wiezorrek, Woillez,
  Yazici, \& Ziegler}]{GravityCollaboration2020}
{GravityCollaboration}, Nowak, M., Lacour, S., {et~al.} 2020, A\&A, 633, A110

\bibitem[{Lacour {et~al.}(2016)Lacour, Biller, Cheetham, Greenbaum, Pearce,
  Marino, Tuthill, Pueyo, Mamajek, Girard, Sivaramakrishnan, Bonnefoy, Baraffe,
  Chauvin, Olofsson, Juhasz, Benisty, Pott, {Sicilia-Aguilar}, Henning,
  Cardwell, Goodsell, Graham, Hibon, Ingraham, Konopacky, Macintosh,
  Oppenheimer, Perrin, Rantakyr{\"o}, Sadakuni, \& Thomas}]{Lacour2016}
Lacour, S., Biller, B., Cheetham, A., {et~al.} 2016, A\&A, 590, A90

\bibitem[{Lacour {et~al.}(2019)Lacour, Dembet, Abuter, F{\'e}dou, Perrin,
  Choquet, Pfuhl, Eisenhauer, Woillez, Cassaing, Wieprecht, Ott, Wiezorrek,
  Tristram, Wolff, Ram{\'i}rez, Haubois, Perraut, Straubmeier, Brandner, \&
  Amorim}]{Lacour2019}
Lacour, S., Dembet, R., Abuter, R., {et~al.} 2019, A\&A, 624, A99

\bibitem[{Lapeyrere {et~al.}(2014)Lapeyrere, Kervella, Lacour, Azouaoui,
  {Garcia-Dabo}, Perrin, Eisenhauer, Perraut, Straubmeier, Amorim, \&
  Brandner}]{Lapeyrere2014}
Lapeyrere, V., Kervella, P., Lacour, S., {et~al.} 2014, in Optical and
  {{Infrared Interferometry IV}}, Vol. 9146 ({SPIE}), 735--743

\bibitem[{{Lubow} {et~al.}(2015){Lubow}, {Martin}, \& {Nixon}}]{2015Lubow}
{Lubow}, S.~H., {Martin}, R.~G., \& {Nixon}, C. 2015, \apj, 800, 96

\bibitem[{Marino {et~al.}(2015)Marino, Perez, \& Casassus}]{Marino2015}
Marino, S., Perez, S., \& Casassus, S. 2015, ApJL, 798, L44

\bibitem[{Mendigut{\'i}a {et~al.}(2014)Mendigut{\'i}a, Fairlamb, Montesinos,
  Oudmaijer, Najita, Brittain, \& van~den Ancker}]{Mendigutia2014}
Mendigut{\'i}a, I., Fairlamb, J., Montesinos, B., {et~al.} 2014, ApJ, 790, 21

\bibitem[{{Nealon} {et~al.}(2015){Nealon}, {Price}, \& {Nixon}}]{2015Nealon}
{Nealon}, R., {Price}, D.~J., \& {Nixon}, C.~J. 2015, \mnras, 448, 1526

\bibitem[{Perez {et~al.}(2014)Perez, Casassus, M{\'e}nard, Roman, van~der Plas,
  Cieza, Pinte, Christiaens, \& Hales}]{Perez2014}
Perez, S., Casassus, S., M{\'e}nard, F., {et~al.} 2014, ApJ, 798, 85

\bibitem[{{Pinte} {et~al.}(2009){Pinte}, {Harries}, {Min}, {Watson},
  {Dullemond}, {Woitke}, {M{\'e}nard}, \& {Dur{\'a}n-Rojas}}]{2009Pinte}
{Pinte}, C., {Harries}, T.~J., {Min}, M., {et~al.} 2009, \aap, 498, 967

\bibitem[{{Pinte} {et~al.}(2006){Pinte}, {M{\'e}nard}, {Duch{\^e}ne}, \&
  {Bastien}}]{2006Pinte}
{Pinte}, C., {M{\'e}nard}, F., {Duch{\^e}ne}, G., \& {Bastien}, P. 2006, \aap,
  459, 797

\bibitem[{Pinte {et~al.}(2018)Pinte, Price, Menard, Duchene, Dent, Hill, {de
  Gregorio-Monsalvo}, Hales, \& Mentiplay}]{Pinte2018}
Pinte, C., Price, D.~J., Menard, F., {et~al.} 2018, ApJ, 860, L13

\bibitem[{Price {et~al.}(2018{\natexlab{a}})Price, Cuello, Pinte, Mentiplay,
  Casassus, Christiaens, Kennedy, Cuadra, Sebastian~Perez, Marino, Armitage,
  Zurlo, Juhasz, Ragusa, Laibe, \& Lodato}]{Price2018}
Price, D.~J., Cuello, N., Pinte, C., {et~al.} 2018{\natexlab{a}}, Monthly
  Notices of the Royal Astronomical Society, 477, 1270

\bibitem[{Price {et~al.}(2018{\natexlab{b}})Price, Wurster, Tricco, Nixon,
  Toupin, Pettitt, Chan, Mentiplay, Laibe, Glover, Dobbs, Nealon, Liptai,
  Worpel, Bonnerot, Dipierro, Ballabio, Ragusa, Federrath, Iaconi, Reichardt,
  Forgan, Hutchison, Constantino, Ayliffe, Hirsh, \& Lodato}]{Price2018a}
Price, D.~J., Wurster, J., Tricco, T.~S., {et~al.} 2018{\natexlab{b}},
  Publications of the Astronomical Society of Australia, 35

\bibitem[{{van Boekel} {et~al.}(2004){van Boekel}, Min, Leinert, Waters,
  Richichi, Chesneau, Dominik, Jaffe, Dutrey, Graser, Henning, {de Jong},
  K{\"o}hler, {de Koter}, Lopez, Malbet, Morel, Paresce, Perrin, Preibisch,
  Przygodda, Sch{\"o}ller, \& Wittkowski}]{vanBoekel2004}
{van Boekel}, R., Min, M., Leinert, C., {et~al.} 2004, Nature, 432, 479

\bibitem[{Verhoeff {et~al.}(2011)Verhoeff, Min, Pantin, Waters, Tielens, Honda,
  Fujiwara, Bouwman, van Boekel, Dougherty, de~Koter, Dominik, \&
  Mulders}]{Verhoeff2011}
Verhoeff, A.~P., Min, M., Pantin, E., {et~al.} 2011, A\&A, 528, A91

\bibitem[{Vousden {et~al.}(2016)Vousden, Farr, \& Mandel}]{Vousden2016}
Vousden, W.~D., Farr, W.~M., \& Mandel, I. 2016, Monthly Notices of the Royal
  Astronomical Society, 455, 1919

\end{thebibliography}

\end{document}